\documentclass[12pt]{article}
\pdfoutput=1
\usepackage[T1]{fontenc}
\usepackage[utf8]{inputenc}
\usepackage{slashbox}
\usepackage{amssymb}
\usepackage{amsmath,amsthm}
\usepackage{bbm,latexsym}
\usepackage{epsfig}
\usepackage[numbers,sort&compress]{natbib}
\usepackage{hyperref}

\usepackage{graphicx}

\def\L{\mathcal L}
\def\e{\varepsilon}

\newcommand{\wt}{\widetilde}

\begin{document}

\def\a{\alpha}
\def\b{\beta}
\def\c{\chi}
\def\d{\delta}
\def\e{\epsilon}
\def\f{\phi}
\def\g{\gamma}
\def\h{\eta}
\def\i{\iota}
\def\j{\psi}
\def\k{\kappa}
\def\l{\lambda}
\def\m{\mu}
\def\n{\nu}
\def\o{\omega}
\def\p{\pi}
\def\q{\theta}
\def\r{\rho}
\def\s{\sigma}
\def\t{\tau}
\def\u{\upsilon}
\def\x{\xi}
\def\z{\zeta}
\def\D{\Delta}
\def\F{\Phi}
\def\G{\Gamma}
\def\J{\Psi}
\def\L{\Lambda}
\def\O{\Omega}
\def\P{\Pi}
\def\Q{\Theta}
\def\S{\Sigma}
\def\U{\Upsilon}
\def\X{\Xi}

\def\ve{\varepsilon}
\def\vf{\varphi}
\def\vr{\varrho}
\def\vs{\varsigma}
\def\vq{\vartheta}

\def\dg{\dagger}                                
\def\ddg{\ddagger}                            
\def\wt#1{\widetilde{#1}}                    
\def\mt{\widetilde{m}_1}
\def\mti{\widetilde{m}_i}
\def\rt{\widetilde{r}_1}
\def\mtt{\widetilde{m}_2}
\def\mttt{\widetilde{m}_3}
\def\rtt{\widetilde{r}_2}
\def\mb{\overline{m}}
\def\VEV#1{\left\langle #1\right\rangle}        
\def\be{\begin{equation}}
\def\ee{\end{equation}}
\def\ds{\displaystyle}
\def\ra{\rightarrow}

\def\bea{\begin{eqnarray}}
\def\eea{\end{eqnarray}}
\def\NO{\nonumber}
\def\Bar#1{\overline{#1}}


\def\pl#1#2#3{Phys.~Lett.~{\bf B {#1}} ({#2}) #3}
\def\np#1#2#3{Nucl.~Phys.~{\bf B {#1}} ({#2}) #3}
\def\prl#1#2#3{Phys.~Rev.~Lett.~{\bf #1} ({#2}) #3}
\def\pr#1#2#3{Phys.~Rev.~{\bf D {#1}} ({#2}) #3}
\def\zp#1#2#3{Z.~Phys.~{\bf C {#1}} ({#2}) #3}
\def\cqg#1#2#3{Class.~and Quantum Grav.~{\bf {#1}} ({#2}) #3}
\def\cmp#1#2#3{Commun.~Math.~Phys.~{\bf {#1}} ({#2}) #3}
\def\jmp#1#2#3{J.~Math.~Phys.~{\bf {#1}} ({#2}) #3}
\def\ap#1#2#3{Ann.~of Phys.~{\bf {#1}} ({#2}) #3}
\def\prep#1#2#3{Phys.~Rep.~{\bf {#1}C} ({#2}) #3}
\def\ptp#1#2#3{Progr.~Theor.~Phys.~{\bf {#1}} ({#2}) #3}
\def\ijmp#1#2#3{Int.~J.~Mod.~Phys.~{\bf A {#1}} ({#2}) #3}
\def\mpl#1#2#3{Mod.~Phys.~Lett.~{\bf A {#1}} ({#2}) #3}
\def\nc#1#2#3{Nuovo Cim.~{\bf {#1}} ({#2}) #3}
\def\ibid#1#2#3{{\it ibid.}~{\bf {#1}} ({#2}) #3}

\title{
{\normalsize \mbox{ }\hfill
\begin{minipage}{3cm}
IFIC/16-42
\end{minipage}}\\
\vspace*{1mm}
\bf\large Unifying leptogenesis, dark matter and high-energy neutrinos with right-handed neutrino mixing via Higgs~portal\footnote{In this version (v3 on the arXiv), we corrected the calculation of the $N_S$ abundance
in the hierarchical case at the resonance (see Eq.~(\ref{NSzreshier})), including
a suppression factor $M_S/M_{DM}$ missed in previous versions. This makes
the upper bound on $M_{DM}$ more stringent. If also successful leptogenesis is required, 
the case of a PeV dark matter mass is now ruled out 
but the case of a  $\sim 100\,{\rm TeV}$ dark matter masses is still allowed.}
}
\author{
{\large Pasquale Di Bari$^1$, Patrick Otto Ludl$^1$ and Sergio Palomares-Ruiz$^2$}
\\
$^1${\it Physics and Astronomy}, 
{\it University of Southampton,} 
{\it  Southampton, SO17 1BJ, U.K.} \\
$^2${\it Instituto de Física Corpuscular (IFIC)}, 
{\it CSIC-Universitat de València},\\
{\it Apartado de Correos 22085, E-46071 Valencia, Spain}
}

\maketitle \thispagestyle{empty}

\vspace{-9mm}

\begin{abstract}
We revisit a model in which neutrino masses and mixing are described by a two right-handed (RH) neutrino seesaw scenario, implying a strictly hierarchical light neutrino spectrum. A third decoupled RH neutrino, $N_{\rm DM}$ with mass $M_{\rm DM}$, plays the role of cold dark matter (DM) and is produced by the mixing with a source RH neutrino, $N_{\rm S}$ with mass $M_{\rm S}$, induced by Higgs portal interactions. The same interactions are also responsible for $N_{\rm DM}$ decays. We discuss in detail the constraints coming from DM abundance and stability conditions showing that in the hierarchical case, for $M_{\rm DM}\gg M_{\rm S}$, there is an allowed window,  $100\,{\rm GeV} \lesssim M_{\rm DM} \lesssim 10\,{\rm PeV}$, necessarily implying a contribution, from DM decays, to the high-energy neutrino flux recently detected by IceCube. We also show how the decays of the two coupled RH neutrinos can explain the matter-antimatter asymmetry of the universe via leptogenesis in the quasi-degenerate limit. In this case, the DM mass should be within a tighter range 
300~GeV $\lesssim M_{\rm S} < M_{\rm DM} \lesssim$ 300~TeV. We discuss the specific properties of this high-energy neutrino flux and show the predicted event spectrum for two exemplary cases. Although DM decays, with a relatively hard spectrum, cannot account for all the IceCube high-energy data, we illustrate how this extra source of high-energy neutrinos could reasonably explain some potential features in the observed spectrum. In this way, this represents a unified scenario for leptogenesis and DM that could be tested during the next years with more high-energy neutrino events. 

\end{abstract}
\newpage
\section{Introduction}

The possibility of explaining the dark matter (DM) and baryon asymmetry of the universe within a unified picture is an attractive idea, intensively explored during recent years~\cite{reviews}. This is also motivated by the simple observation that baryons and DM give a similar contribution to the cosmic energy budget.

Moreover, since both DM and baryon asymmetry of the universe require new physics, it is conceivable that they should be ultimately explained within a common model. An extension of the standard model (SM) is also required by neutrino masses and mixing established by neutrino oscillation experiments and, therefore, it is quite natural to look at neutrino physics as a possible source for cosmological DM and matter-antimatter asymmetry.\footnote{Curiously within errors one has $\O_{\rm DM,0}/\O_{\rm B,0} \simeq m_{\rm atm}/m_{\rm sol} \simeq 5$, where $m_{\rm atm}$ and $m_{\rm sol}$ are the atmospheric and solar neutrino mass scales, respectively.} The simplest way to describe neutrino masses and mixing is by adding to the SM Lagrangian right-handed (RH) neutrino Yukawa couplings and a Majorana mass term. In the seesaw limit one obtains the seesaw formula for the low-energy neutrino mass, nicely explaining why left-handed (LH) neutrinos are much lighter compared to all other massive fermions~\cite{seesaw}. 

It is then natural to think whether heavy RH neutrinos can play a cosmological role. In the traditional high-energy scale leptogenesis scenario~\cite{fy}, RH neutrino decays are the source of the observed baryon asymmetry. In the $\nu$MSM scenario~\cite{nuMSM}, the lightest RH neutrino with ${\cal O}$(keV) mass can play the role of DM, with the correct abundance produced by active-sterile neutrino mixing. At the same time, the mixing between the two heavier RH neutrinos can also produce the observed baryon asymmetry~\cite{ars}. In this model, however, the neutrino Yukawa couplings are many orders of magnitude smaller than those of all other particles and somehow the original motivation of the seesaw is not addressed.\footnote{However, in Ref.~\cite{shapmodel} it was shown how the specific $\nu$MSM neutrino Yukawa matrix could arise as a consequence of a lepton number symmetry slightly broken by the Majorana mass terms and Yukawa coupling constants. For more details on the $\nu$MSM model we refer the reader to Ref.~\cite{gorbunov}.} 

As a remedy, it was proposed~\cite{CDMRHnumix} that if one of the RH neutrinos decouples and it is stable on cosmological time scales, it could play the role of DM and the same mixing among RH neutrinos could also reproduce the correct DM abundance. This is possible if one introduces a non-renormalizable operator $\l_{AB}\,\phi^{\dagger}\,\phi\,\overline{N^c_A} \, N_B/\L$, a simple example of Higgs portal models~\cite{wilczek}, arising from new physics at some scale $\L$ with additional couplings $\l_{AB}$.  The presence of this new interaction would enhance medium effects opening a new production mechanism from RH-RH neutrino mixing (occurring much above the electroweak scale) rather than from RH-LH neutrino mixing (occurring much below the electroweak scale), as in the $\nu$MSM. At the same time it would also be responsible for the DM RH neutrino, $N_{\rm DM}$ with mass $M_{\rm DM}$, eventually to decay, producing high-energy neutrinos\footnote{For limits on DM decays with neutrino detectors, see Refs.~\cite{PalomaresRuiz:2007ry, Covi:2009xn, Esmaili:2012us, Murase:2012xs}.} and it was pointed out\footnote{In Ref.~\cite{CDMRHnumix} it was noticed that a possible explanation of the PAMELA excess would also imply a potential signal at the IceCube detector.} that this flux could be detectable at neutrino telescopes such as IceCube~\cite{CDMRHnumix}.

Intriguingly, the IceCube detector has recently detected the first ever high-energy neutrino events of extraterrestrial origin~\cite{icecube, icecube2}, i.e., that cannot be accounted for by the known atmospheric neutrino flux. The energies of these events are as high as ${\cal O}({\rm PeV})$, i.e., within the natural range of masses needed by the mechanism of cold DM from RH neutrino mixing, as we will discuss in detail. Indeed, very heavy DM decays have been proposed to account for part or all of these events~\cite{Feldstein:2013kka, serpico} and different constraints, within different models, have been presented~\cite{Bai:2013nga, Bhattacharya:2014vwa, Rott:2014kfa, Esmaili:2014rma, Fong:2014bsa, murase, Esmaili:2015xpa, Boucenna:2015tra, Daikoku:2015vsa, Chianese:2016opp, Fiorentin:2016avj, Dev:2016qbd}, showing that this could be a potential explanation of the observed events (or part of them). Encouraged by this phenomenological picture, in this paper we revisit the cold DM RH neutrino mixing scenario discussing a few important aspects and showing how to test it with high-energy neutrino detectors such as IceCube and also showing explicitly how the same set up can accommodate the matter-antimatter asymmetry of the universe via leptogenesis, as first pointed out\footnote{A cosmologically stable RH neutrino playing the role of DM has also been proposed within left-right symmetric models~\cite{leftright1}, also in combination with leptogenesis~\cite{leftright2}, and within a hybrid seesaw model~\cite{hybrid}.} in Refs.~\cite{CDMRHnumix, anisimov, unpublished}.  

The paper is organised as follows. In Section~\ref{sec:scenario} we review the idea of the seesaw mechanism with one decoupled RH neutrino and the cold DM RH neutrino mixing scenario. In particular we  discuss in detail how in the simplest scenario, in order to reproduce the correct DM abundance with a cosmologically stable candidate, $M_{\rm DM}$ has to be in the range $100$~GeV--$1$~EeV in the hierarchical case, for $M_{\rm DM}\gg M_{\rm S}$.  We also discuss a few ingredients, some of which represent quite plausible possibilities, that might relax the constraints especially in the quasi-degenerate limit with $M_{\rm S}\simeq M_{\rm DM}$. In Section~\ref{sec:asymmetry} we show how the observed matter-antimatter asymmetry can be explained in this model via leptogenesis and in this case the range for $M_{\rm DM}$ restricts to $\sim$~TeV--10~PeV intriguingly overlapping with the range of neutrino energies detected at IceCube. In Section~\ref{sec:IceCube} we discuss the features of the predicted high-energy neutrino flux and its related event spectrum in the IceCube detector, for two exemplary cases for which the DM mass is above a few 100~TeV. We compare these results with the observed event spectrum and show that, although a DM-only signal cannot explain the 4-year IceCube spectrum, it could help to explain some of its features. Finally, in Section~\ref{sec:conclusions} we draw our conclusions.

\section{The cold DM RH neutrino mixing scenario}
\label{sec:scenario}

In this section, we revisit the DM RH neutrino scenario~\cite{CDMRHnumix}, combine all constraints using the most recent data and we finally discuss some plausible viable scenarios and the allowed range of values for  $M_{\rm DM}$. In particular we show that relaxing the assumption of ultra-relativistic thermal $N_{\rm S}$ abundance at the resonance, the hierarchical case ($M_{\rm DM} \gg M_{\rm S}$) becomes viable extending the range of allowed values for $M_{\rm DM}$.

\subsection{From the minimal seesaw Lagrangian to Higgs portal interactions}

We assume the usual minimal SM extension with three RH neutrinos $N_i$ with Yukawa couplings $h$ and a Majorana mass matrix $M$. After spontaneous symmetry breaking, the Higgs vacuum expectation value (vev) $v$ generates a neutrino Dirac mass so that the neutrino mass terms can be written as ($\a = e, \m, \t; \, i=1,2,3$)
\be\label{M+D}
- {\cal L}_{M} =  \overline{\nu_{\a L}}\,m_{D\a i} \, N_{i R} +
                          \frac{1}{2} \, \overline{N^{c}_{i R}} \, D_{Mii} \, N_{i R}  + \mbox{\rm h.c.} ~,
\ee
where  $D_M \equiv {\rm diag}(M_1, M_2, M_3)$ with $M_1 \leq  M_2 \leq M_3$, in a basis where the Majorana and charged lepton mass matrices are diagonal. In the seesaw limit, $D_M\gg m_D$, the set of neutrino masses splits into a light neutrino set with masses given by the seesaw formula
\be\label{seesaw}
D_m \equiv {\rm diag}(m_1,m_2,m_3) =  U^{\dagger} \, m_D \, \frac{1}{D_M} \, m_D^T  \, U^{\star} ~,
\ee
where $U$ is the leptonic mixing matrix and for the light neutrino masses we adopt the convention $m_1 \leq m_2 \leq m_3$. From neutrino oscillation experiment global analyses we know that~\cite{global,nufit} $m_{\rm sol}^2 = (7.49^{+0.19}_{-0.17}) \times 10^{-5} \, {\rm eV}^2$, where $m^2_{\rm sol} \equiv m^2_2 -m^2_1$ for normal ordering (NO) and $m_{\rm sol}^2 \equiv m^2_3 -m^2_2$ for inverted ordering (IO), and $m_{\rm atm}^2 = (2.477^{+0.042}_{-0.042}) \times 10^{-5} \, {\rm eV}^2$ for NO and $m_{\rm atm}^2  = (2.465^{+0.041}_{-0.043}) \times 10^{-5} \, {\rm eV}^2$ for IO, where $m^2_{\rm atm}\equiv m^2_3 -m^2_1$.

We parameterise the mixing matrix $U$ in the usual way for NO,
\begin{equation}
U^{(\rm NO)} = 
\left(
\begin{array}{ccc}
c_{13} c_{12} &
c_{13} s_{12} &
s_{13} e^{-i \delta} \\
- c_{23} s_{12} - s_{23} s_{13} c_{12} e^{i \delta} &
c_{23} c_{12} - s_{23} s_{13} s_{12} e^{i \delta} &
s_{23} c_{13} \\
s_{23} s_{12} - c_{23} s_{13} c_{12} e^{i \delta} &
-s_{23} c_{12} - c_{23} s_{13} s_{12} e^{i \delta} &
c_{23} c_{13}
\end{array}
\right) \, \mathrm{diag}(e^{i\rho},\, e^{i\sigma},\, 1) ~,
\end{equation}
where $c_{ij} = \cos \theta_{ij}$, $s_{ij} = \mathrm{sin}\theta_{ij}$, $\delta$ is the Dirac phase and $\rho$ and $\sigma$ are the two Majorana phases. For IO, since we use the convention $m_1 < m_2 < m_3$, this has to be cyclically permuted such that
\be
U^{(\rm IO)} = U^{(\rm NO)}\, 
\begin{pmatrix}
0 & 1 & 0\\
0 & 0 & 1\\
1 & 0 & 0  
\end{pmatrix} ~.
\ee
The best fit values and $1\s$ ($3\s$) confidence level (C.L.) ranges of the reactor, solar and atmospheric mixing angles for NO (left column) and IO (right column), are given by~\cite{global,nufit}
\bea\label{eq:expranges}
\nonumber
\theta_{13} & = &  {8.50^{\circ}}^{+0.19^{\circ}}_{-0.20^{\circ}} \, \;\;  
(7.87^{\circ}\mbox{--}9.08^{\circ}) 
 \hspace{9mm} \mbox{\rm and} \hspace{9mm}
 \theta_{13}  =    {8.51^{\circ}}^{+0.20^{\circ}}_{-0.20^{\circ}} \, \;\; 
 (7.89^{\circ}\mbox{--}9.10^{\circ}) \,  ,
 \\ \nonumber
\theta_{12} & = &  {33.72^{\circ}}^{+0.79^{\circ}}_{-0.76^{\circ}}\,  \;\;  
(31.52^{\circ}\mbox{--}36.18^{\circ}) 
\hspace{3mm} \mbox{\rm and} \hspace{9mm}
\theta_{12}  =   {33.72^{\circ}}^{+0.79^{\circ}}_{-0.76^{\circ}}\,  \;\;  
(31.52^{\circ}\mbox{--}36.18^{\circ}) \,  , 
\\
\theta_{23} & = &  {42.2^{\circ}}^{+2.2^{\circ}}_{-1.4^{\circ}} \,  \;\;  
(38.5^{\circ}\mbox{--}52.8^{\circ}) 
\hspace{11mm} \mbox{\rm and} \hspace{9mm}
\theta_{23}  =   
                 {49.4^{\circ}}^{+1.4^{\circ}}_{-1.9^{\circ}}  
                  \;\;  (38.8^{\circ}\mbox{--}52.9^{\circ})  ~,
\eea 
where the LID method of the NO$\nu$A $\nu_e$-appearance data has been considered. In addition, current experimental data also start to set constraints on the Dirac phase and the best fit values and $1\s$ C.L. intervals are found to be, for NO and IO respectively, 
\be\label{delta}
\d  = {303^{\circ}}^{+39^{\circ}}_{-50^{\circ}} \hspace{5mm}  
\;\;\mbox{\rm and} \hspace{5mm}
\d  = {262^{\circ}}^{+51^{\circ}}_{-57^{\circ}}  ~,
\ee
though all values $[0^{\circ}, 360^{\circ}]$ are still allowed at $3\s$ C.L.\ for NO and $[0^\circ,56^\circ] \cup [98^\circ, 360^\circ]$ for IO. On the other hand, although the best fit in this analysis is obtained for IO, none of the two orderings is clearly favoured over the other (using different methods gives rise to different results in this respect~\cite{global,nufit}). 

We assume that one of the three RH neutrinos, $N_{\rm DM}$, has Yukawa couplings small enough to guarantee its stability on cosmological time scales so that it is a potential DM candidate~\cite{CDMRHnumix}. In this case, necessarily, the neutrino Dirac mass matrix has to be written in one of the three following forms, 
\be\label{WASS}
 m_D \simeq
 \left( \begin{array}{ccc}
0 &  m_{D e 2}  &  m_{D e 3}  \\
0 &  m_{D \m 2}  &  m_{D \m 3} \\
0 &  m_{D \t 2}  &  m_{D \t 3}
\end{array}\right) \,     , \,  {\rm or} \,
\left( \begin{array}{ccc}
m_{D e 1} & 0  &  m_{D e 3}  \\
m_{D \m 1} & 0 &  m_{D \m 3} \\
m_{D \t 1}  & 0 &  m_{D \t 3}
\end{array}\right) \,   , \, {\rm or} \, 
\left( \begin{array}{ccc}
m_{D e 1}    &  m_{D e 2}  & 0  \\
m_{D \m 1} &  m_{D \m 2} & 0 \\
m_{D \t 1}  &  m_{D \t 2} & 0
\end{array}\right) \,   ,
\ee
corresponding effectively to a two-RH neutrino model in which either the lightest RH neutrino $N_1$, the next-to-lightest $N_2$ or the heaviest $N_3$ is decoupled and has to be identified with $N_{\rm DM}$.\footnote{These special $m_D$ forms  can be easily justified imposing for example a $Z_2$ symmetry under which $N_{\rm DM}$ is odd and the other two RH neutrinos are even. The same forms have also been considered in a different context in order to have resonant leptogenesis testable at colliders~\cite{teresi}.} These three forms for $m_D$, with three texture zeros, can be parameterised in terms of nine physical parameters, since three phases can be always reabsorbed in the LH neutrino fields. In the Yukawa basis the Dirac mass matrix is diagonal and we can express it as $D_{m_D} \equiv v\, {\rm diag}(h_{A}, h_{B}, h_{C})$, with $h_{A} \leq h_{B} \leq h_{C}$. The transformation from the basis where the charged lepton and Majorana mass matrices are diagonal to the Yukawa basis can be described in terms of two unitary matrices, $V_L$ and $U_R$, acting respectively on the LH and on the RH neutrinos, explicitly
\be\label{biunitary}
m_D = V_L^{\dagger}\, D_{m_D} \, U_R  \,  .
\ee
Our working assumption, Eq.~(\ref{WASS}), necessarily implies $h_{A} \simeq 0$ and consequently, from the seesaw formula, Eq.~(\ref{seesaw}), one has $m_1 \simeq 0$: in our scenario light neutrinos are strictly hierarchical, either normal hierarchy (NH) or inverted hierarchy (IH). 
The matrix $U_R$ is the RH neutrino mixing matrix and connects the Yukawa eigenstates $N_J$ to the mass eigenstates $N_k$: $N_J = (U_{R})_{Jk}\,N_k$ ($J=A,B,C$). It can be regarded as the analogue of the leptonic mixing matrix for the light neutrinos and it can be similarly parameterised by three mixing angles $\theta_{ij}^R$ and three phases. If $h_A=0$, then $N_{\rm DM}$ is strictly stable and it coincides exactly with $N_A$, implying that the two mixing angles of $N_{\rm DM}$ with the other two RH neutrinos vanish.

We can also conveniently express $m_D$ in the orthogonal parameterisation~\cite{casasibarra},
\be\label{orthogonal1}
m_D=U\,\sqrt{D_m}\,\O\,\sqrt{D_M} \,  ,
\ee
where $\O$ is the orthogonal matrix encoding information on the RH neutrino total decay widths and total $C\!P$ asymmetries. The three forms for $m_D$ in Eq.~(\ref{WASS}) necessarily imply, respectively, the three following forms for $\O$: 
\be\label{orthogonal}
\O \simeq  \left(
\begin{array}{ccc}
  1  &  0   & 0   \\
  0  & \cos \o             &  \sin \o \\
  0 &  -\zeta\sin \o & \zeta\cos \o
\end{array}
\right) \,  ,  \, {\rm or} \,
\left(
\begin{array}{ccc}
 0   &  1   &  0   \\
  \cos \o  &      0        & \sin \o \\
  -\z\sin \o &  0 & \z \cos \o
\end{array}
\right) \, ,  \, {\rm or} \,
\left(
\begin{array}{ccc}
   0  &  0    & 1  \\
    \cos \o  &   \sin \o     & 0  \\
 -\z \sin \o & \z \cos \o & 0 
\end{array}
\right)   \,  ,
\ee
where $\o$ is a complex angle and $\z = \pm 1$ is a discrete parameter and the two possible values correspond to two different distinct branches of $\Omega$, with positive and negative determinant respectively~\cite{2RHN}. Notice that the nine (real) parameters needed to parameterise the Dirac mass matrix are in this case given by five parameters in the leptonic mixing matrix $U$ (three mixing angles, one Dirac phase, one Majorana phase), two LH neutrino masses, $m_2$ and $m_3$, and finally two real parameters in the complex mixing angle $\o$.

If Eq.~(\ref{WASS}) is assumed to hold exactly, then $N_{\rm DM}$ would be strictly stable. However, in this case $N_{\rm DM}$ could not be produced by any interaction, except maybe via gravitational ones, for example at the end of inflation, if it has a mass close to the inflaton mass~\cite{wimpzilla}. In that case, $N_{\rm DM}$ would likely be identified with the heaviest RH neutrino, $N_3$. At the same time, it would be questionable whether such a particle would exist at all, not having any interaction. One could think of solving both problems by perturbing the form Eq.~(\ref{WASS}) for the Dirac mass matrix, introducing some tiny Yukawa coupling $h_{A}$. In this case $N_{\rm DM}$ would decay with a lifetime (after EW symmetry breaking)
\be
\tau_{\rm DM} = \frac{4 \, \pi}{h^2_{A}\,M_{\rm DM}} \simeq 0.87 \, h^{-2}_A \, 10^{-23} \, 
\left(\frac{{\rm GeV}}{M_{\rm DM}}\right) \,{\rm s} ~.
\ee
From the latest IceCube results, as we will discuss in detail, one has to require $\tau_{\rm DM} > \t_{\rm DM}^{\rm min}\simeq 10^{28}\,{\rm s}$, so that the DM Yukawa coupling would be
\be
h_{A}  \lesssim 3 \times 10^{-26} \, \sqrt{\frac{\rm GeV}{M_{\rm DM}}} \,  ,
\ee
which is too tiny to think of any DM production mechanism via Yukawa interactions. One possible solution, the so called $\nu$MSM model~\cite{nuMSM}, is to have the lightest RH neutrino sufficiently light to dominantly decay into three ordinary neutrinos,  in a way that $\t_1 \propto M_1^5$. At the same time, in this way the LH-RH neutrino mixing angle is enhanced and this would induce a sizeable RH neutrino production from mixing. The conditions for the cosmological stability and the correct DM abundance can then be satisfied for a mass of the lightest RH neutrino in the keV range.

However, this solution has the disadvantage of a drastic suppression of all three Yukawa couplings compared to those of all other massive fermions. An alternative possibility is to produce $N_{\rm DM}$, not necessarily the lightest, through the mixing with the other (two) thermalised RH neutrinos~\cite{CDMRHnumix}. However, in this case it is easy to see that the mixing cannot be the minimal mixing encoded in the $U_R$ matrix, defined by Eq.~(\ref{biunitary}), which describes the mismatch between the Yukawa basis and the basis where the RH neutrino mass matrix is diagonal. This is so since the mixing angles of $N_{\rm DM}$ with the other two RH neutrinos $N_j$ would be $\theta_{Aj} \lesssim h_A/h_{\rm I}$ ($I=B,C$), too tiny to produce a sizeable $N_{\rm DM}$ abundance, as $\O_{\rm DM}\,h^2 \propto \theta^2_{Aj}$. For all practical purposes, we can then neglect such small mixing angles and consider $h_A=0$.

A way out is to introduce non-standard RH neutrino interactions originating from new physics at an effective scale $\L$.  At lower energies, this gives rise to a non-renormalizable effective operator, an example of Higgs portal interactions~\cite{wilczek}, which in the Yukawa basis can be written as\footnote{Here, we do not refer to any specific model generating this interaction and, therefore, we will treat $\L$ as a free phenomenological parameter. Below, we will give an example of a simple model able to justify the large values of $\L$ that we will obtain.} ($I,J=A,B,C$)~\cite{anisimov, CDMRHnumix, bezrukov}
\be\label{newoperator}
{\cal L} = \frac{\l_{IJ}}{\L}\, \phi^{\dagger} \, \phi \, \overline{N^c_I}\,N_J \,   .
\ee
Let us show that when introducing this new non-standard interaction, the RH neutrino mixing can lead to the correct $N_{\rm DM}$ abundance.

\subsection{Estimation of the $N_{\rm DM}$ abundance}

In general, the new interaction couplings are non-diagonal in the Yukawa basis and this can provide an efficient source for the RH neutrino mixing. Indeed they would give a contribution to the effective matter potential of the RH neutrino, which in the Yukawa basis is given by~\cite{CDMRHnumix}
\be\label{LAMBDAmixing}
V^{\L}_{JK} \simeq \frac{T^2}{12\,\L}\,\l_{JK} \,  \hspace{3mm} (J,K=A,B,C) .
\ee
On the other hand, the Yukawa interactions clearly produce a diagonal contribution to the RH neutrino Hamiltonian in the Yukawa basis given by~\cite{weldon}
\be
V^{Y}_J =  \frac{T^2}{8\,E_J} \, h^2_J \, \hspace{3mm} (J=A,B,C)  ,
\ee
where $E_J$ is the $N_J$ energy and RH neutrinos are assumed to be in ultra-relativistic thermal equilibrium (of course, this is not true for $N_{\rm DM}=N_A$, since in this case $h_A =0$). In the mass eigenstates basis one also has the usual kinetic contribution.

Let us assume that the mass eigenstate corresponding to $N_{\rm DM}$ mixes with just one of the other two thermalised (mass eigenstate) RH neutrinos. This would play the role of the {\em source} RH neutrino, that we refer to as $N_{\rm S}$, and which is in general a linear combination of $N_B$ and $N_C$.
 
In this way, we have a simple two-neutrino mixing formalism, neglecting for the time being the mixing with the third RH neutrino mass eigenstate $N_{\rm I}$ (the {\em interfering} RH neutrino). We will comment on this at the end of this section, showing that indeed the mixing with this RH neutrino can be neglected, since only the mixing with one RH neutrino can satisfy simultaneously all constraints. However, in Section 3 we will see that $N_{\rm I}$ plays a crucial role for leptogenesis (and of course, it is in any case necessary in order to reproduce correctly the neutrino oscillation experimental data).

The source RH neutrino has a Yukawa coupling $h_{\rm S} \equiv \sqrt{(h^{\dagger}\,h)_{ii}}\; ,$ where $i$ is the index corresponding to the mass eigenstate coinciding with $N_{\rm S}$. If we introduce the effective neutrino mass $\widetilde{m}_{\rm S} \equiv v^2 \, h^2_{\rm S} / M_{\rm S}$ and parametrise it as $\widetilde{m}_{\rm S} \equiv \a_{\rm S} \, m_{\rm sol}$, then we can write\footnote{For example, if we again consider the third form for $m_D$ in Eq.~(\ref{WASS}) corresponding to the third form for $\O$ in Eq.~(\ref{orthogonal}) and identify $N_{\rm S}$ with $N_2$, then the effective neutrino mass $\widetilde{m}_{\rm S}$ can be expressed in terms of $\o$ as
\be
\widetilde{m}_{\rm S} = m_{\rm sol}\, |\sin\o|^2 + 
m_{\rm atm} \, |\cos\o |^2  ~.
\ee
Clearly, one has $\widetilde{m}_{\rm S} \geq m_{\rm sol}$, corresponding to $\a_{\rm S} \geq 1$. There is no upper bound for $\a_{\rm S}$ but values $\a_{\rm S} \gg 1$ correspond to $|\O_{ij}|^2 \gg 1$, necessarily implying a fine-tuning in the seesaw formula at the level of $\sim 1/|\O_{ij}|^2$.}
\be\label{hS}
h_{\rm S} = {\sqrt{\a_{\rm S}\,m_{\rm sol}\,M_{\rm S}}\over v} \simeq  1.7 \times 10^{-8} \, \sqrt{\a_{\rm S}\, \left(\frac{M_{\rm S}}{{\rm GeV}}\right)}   ~ .
\ee
In this way, following the standard procedure, we can write the Hamiltonian for the mixed RH neutrinos in the mass eigenstate basis as
\be\label{hamiltonian}
H =  \left( \begin{array}{cc}
E_{\rm DM} &  \frac{T^2}{12\,\widetilde{\L}} \\[1ex]
\frac{T^2}{12\,\widetilde{\L}}  &  E_{\rm S} + \frac{T^2}{8\,E_{\rm S}} \, h^2_{\rm S}
\end{array}\right)  ~,
\ee
where we defined $\widetilde{\L} \equiv \L/\l_{AS}$ and we (reasonably) assumed that non-standard interactions are much smaller than Yukawa interactions, such that $h^2_{\rm S} \gg 2\,T /\widetilde{\L}$ within the relevant temperature range (see below). In the ultra-relativistic limit we can then write $E_{\rm DM} \simeq p + M^2_{\rm DM}/(2\,p)$ and $E_{\rm S} \simeq p + M^2_{\rm S}/(2\,p)$. As usual, subtracting a contribution to $H$ proportional to the identity that does not contribute to the mixing, we are left with the effective mixing Hamiltonian
\be
\Delta H \simeq   
\left( \begin{array}{cc}
- \frac{\D M^2}{4 \, p} - \frac{T^2}{16\,p} \, h^2_{\rm S} &  \frac{T^2}{12\,\widetilde{\L}}  \\[1ex]
\frac{T^2}{12\,\widetilde{\L}} &  \frac{\D M^2}{4 \, p} + \frac{T^2}{16 \, p} \, h^2_{\rm S}  
\end{array}\right)  ~,
\ee
where we defined $\D M^2 \equiv M^2_{\rm S} - M^2_{\rm DM}$.  If we now describe the neutrino spectrum by its average momentum, $p \simeq 3\,T$, and introduce the dimensionless effective potential $v_{\rm S}^Y \equiv T^2\,h^2_{\rm S} / (4\,\D M^2)$ and the effective mixing angle $\sin 2\theta_{\L}(T) \equiv T^3/(\widetilde{\L} \, \D M^2 )$, due to the presence of the non-standard interactions, we can recast $\D H$ as 
\be
\Delta H \simeq   
\frac{\D M^2}{12\,T}\,\left( \begin{array}{cc}
	- 1 - v_{\rm S}^Y &  \sin 2\theta_{\L}  \\[1ex]
\sin 2\theta_{\L}   &  1 + v_{\rm S}^Y 
\end{array}\right) ~.
\ee
The energy eigenstates in matter have energies $E_{\rm DM}^m(T)$ and $E_{\rm S}^m(T)$. While the temperature drops down, these tend to get closer to the mass eigenstates $N_{\rm DM}$ and $N_{\rm S}$ and
\be
E^{\rm m}_{\rm DM} - E_{\rm S}^{\rm m} \simeq \frac{\D M^2}{6\, T}
\sqrt{\left(1  + v^Y_{\rm S} \right)^2 + \sin^2\,2\theta_{\L}} ~.
\ee
The mixing angle $\theta_m $ is given by
\be
\sin 2\theta_{\L}^{\rm m}= 
\frac{\sin 2\theta_{\L}}{
\sqrt{\left(1  + v^Y_{\rm S} \right)^2 + \sin^2\,2\theta_{\L}}}  ~.
\ee
Notice that also the mixing angle in vacuum $\theta_{\L}$ is a function of the temperature. If $\D M^2 < 0$, equivalent to having $M_{DM} > M_{\rm S}$, there is a resonance for $v_Y = -1$.  This resonance condition is verified for a specific value of the temperature:
\be
T_{\rm res}  \equiv \frac{2 \, \sqrt{|\D M^2|}}{h_{\rm S} } =  \frac{2 \, \sqrt{ M^2_{\rm DM} - M^2_{\rm S}}}{h_{\rm S}} ~.
\ee
It is also useful to introduce the quantity
\be\label{zres}
z_{\rm res} \equiv \frac{M_{\rm DM}}{T_{\rm res}} = \frac{h_{\rm S} \, M_{\rm DM}}{2 \, \sqrt{M^2_{\rm DM}-M^2_{\rm S}}} \simeq  0.85 \times 10^{-8}\,\sqrt{\a_{\rm S}\,\left({M_{\rm S}\over {\rm GeV}}\right)}
\frac{M_{\rm DM}/M_{\rm S}}{\sqrt{M^2_{\rm DM}/M^2_{\rm S}-1}}  ~,
\ee
showing that for a fixed $M_{\rm DM}$ and decreasing $M_{\rm S}$, then $z_{\rm res}$ decreases (i.e. $T_{\rm res}$ increases). This observation will be useful when we will discuss the hierarchical case $M_{\rm DM}/M_{\rm S} \gtrsim 2$.

At $T \gg T_{\rm res}$, the coupled RH neutrino $N^{\L}_{\rm S}(T)$, the interaction eigenstate, is assumed in thermal ultra-relativistic equilibrium (later on we will see that a lower value of the  abundance is also possible and even preferred) and this basically coincides with the matter eigenstate with energy $E_{\rm DM}^{m}$. Since the process is highly non-adiabatic, $N^{\L}_{\rm S}(T)$ does not track the matter eigenstate while the temperature drops down. In this way, at $T_{\rm res}$, just a small fraction of $N^{\L}_{\rm S}(T)$ is converted non-adiabatically into the DM RH neutrinos, $N^{\L}_{\rm DM}(T)$, the interaction eigenstate that in the absence of mixing would coincide with the mass eigenstate $N_{\rm DM}$. Let us stress that even after spontaneous symmetry breaking, at zero temperature, there is still a tiny non vanishing mixing angle, $\theta_{\L}^0$, such that $N^{\L,0}_{\rm DM}$ (the genuine DM state) does not exactly coincide with $N_{\rm DM}$ but also has a tiny $N_{\rm S}$ component, which is one of the reasons why it is not strictly stable as we will see. The fraction of $N^{\L}_{\rm S}$ converted into $N^{\L}_{\rm DM}$ can be calculated using the Landau-Zener formula\footnote{See Ref.~\cite{anisimov} for a derivation within the density matrix formalism.}
\be\label{NDM2NS}
\left.\frac{N_{N_{\rm DM}}}{N_{N_{\rm S}}}\right|_{\rm res} \simeq  \frac{\pi}{2} \, \g_{\rm res}  ~,
\ee
where $\g_{\rm res} $ is the adiabaticity parameter at the resonance, defined as (see, e.g., Ref.~\cite{akhmedov})
\be
\g_{\rm res}  \equiv  \left. \frac{|E^{\rm m}_{\rm DM}-E^{\rm m}_{\rm S}|}{2 \, |\dot{\theta}_m|} \right|_{\rm res}  ~.
\ee
Then, a straightforward calculation gives first
\be
|E^{\rm m}_{\rm DM} - E_{\rm S}^{\rm m}|_{\rm res} \simeq \frac{|\D M^2|}{6\, T_{\rm res}}\,\sin\,2\theta_{\L}  ~,
\ee
\be
\frac{1}{2\,|\dot{\theta}_m|_{\rm res}} = \left. \frac{\sin\,2\theta_{\L}}{|\dot{v}_{\rm S}^Y|}\right|_{\rm res}  ~,
\ee
and then, using $|\dot{v}_Y|_{\rm res} =2\,H_{\rm res}$, we finally find\footnote{Here we correct the factor in the denominator given in Ref.~\cite{CDMRHnumix}, that was $6$ instead of $12$.}
\be\label{gres0}
\g_{\rm res}  = \sin^2 2\theta_{\L}(T_{\rm res}) \, \frac{|\D M^2|}{12\,T_{\rm res}\,H_{\rm res}}  ~,
\ee
where $H_{\rm res} \simeq 1.66\,\sqrt{g^{\rm res}_{\star}}\,T^2_{\rm res}/M_{\rm Pl}$ is the expansion rate at the resonance and $g^{\rm res}_{\star}$ is the number of degrees of freedom at the resonance. This can be assumed to have approximately the SM value plus the contribution from the $N_{\rm S}$, so that $g_{\star}^{\rm res}=g_{\star}^{SM}+7/4=108.5$. In this way we obtain
\be\label{gres}
\g_{\rm res} \simeq 
0.4 \, \frac{M_{\rm Pl} \, \sqrt{|\D M^2|}}{\widetilde{\L}^2 \,\sqrt{g^{\rm res}_{\star}}\, h_{\rm S}^3}  ~.
\ee
For the DM abundance one obtains the simple relation
\be\label{DMabgres}
\O_{\rm DM}\,h^2 \simeq 1.45 \times 10^6\,\left(\frac{N_{\rm DM}}{N_\g}\right)_{\rm res} \, \left(\frac{M_{\rm DM}}{{\rm GeV}}\right)  ~,
\ee
where $(N_{\rm DM}/N_{\g})_{\rm res}$ is the DM-to-photon number ratio at the end of resonant conversion. Since we are assuming that at the resonance the $N_{\rm S}$'s are fully thermalised, then $(N_{S}/N_\g)_{\rm res} \simeq 3/4$ and, using Eq.~(\ref{NDM2NS}), one obtains
\be\label{DMabundance}
\O_{\rm DM}\,h^2 \simeq 1.7 \times 10^6 \, \g_{\rm res} \, \left(\frac{M_{\rm DM}}{{\rm GeV}}\right) ~.
\ee
From Eq.~(\ref{zres}), we can express $\D M^2$ in terms of $z_{\rm res}$,
\be\label{DM2}
\sqrt{|\D M^2|} = \frac{h_{\rm S}\,M_{\rm DM}}{2 \, z_{\rm res}} ~.
\ee
Plugging this expression into Eq.~(\ref{gres}) and re-expressing $h_{\rm S}$ in terms of $\a_{\rm S}$, we find
\be\label{gres2}
\g_{\rm res} \simeq \frac{8}{\a_{\rm S}\,z_{\rm res}}
\, \left(\frac{M_{\rm DM}}{M_{\rm S}}\right) \, \left(\frac{10^{16} \, {\rm GeV}}{\widetilde{\L}}\right)^2  ~.
\ee
In the quasi-degenerate limit, the ratio $M_{\rm DM}/M_{\rm S}$ is simply $M_{\rm DM}/M_{\rm S} \simeq 1$. The assumption that $N_{\rm S}$ is in ultra-relativistic thermal equilibrium at the resonance imposes both a lower bound and an upper bound on $z_{\rm res}$.  The condition that they are ultra-relativistic simply requires $M_{\rm S}/T_\mathrm{res} \lesssim 3$, since for larger values, the $N_{\rm S}$ abundance is Boltzmann suppressed. 

There is also a less trivial lower bound. In our setup, the only interactions that can thermalise $N_{\rm S}$ are the Yukawa couplings.  Assuming that after inflation the $N_{\rm S}$ abundance is negligible\footnote{This is a logical self-consistency condition: if we postulate that some external mechanism produces some amount of $N_{\rm S}$, then the same external mechanism could be invoked to produce also $N_{\rm DM}$. A possibility is that, at the end of inflation, the inflaton field does not couple to $N_{\rm DM}$ but only to $N_{\rm S}$ and in this case, one could even assume an abundance above the thermal value. We will be back to this point at the end of this section.} then $N_{\rm S}$ would thermalise at $z=z_{\rm eq}$, defined as that value of $z$ such that $N_{N_{\rm S}}(z_{\rm eq})=1$ for initially vanishing $N_{\rm S}$ abundance. Since we are assuming $N_{\rm S}$ ultra-relativistic thermal equilibrium, then we have to impose $z_{\rm res} \gtrsim z_{\rm eq}$.

Inverse decays would thermalise $N_{\rm S}$ at $z_{\rm eq} \simeq (6/K_{\rm S})^{1/3} \simeq 0.4\,(100/K_{\rm S})^{1/3}$~\cite{pedestrians}, where $K_{\rm S} \equiv \widetilde{m}_{\rm S}/m_{\star}$. However, when $(2\leftrightarrow 2)$ scatterings involving top quarks and gauge bosons are also taken into account, the $N_{\rm S}$ thermalisation is more efficient and $z_{\rm eq} \simeq 8/K_{\rm S}$, valid for $K_{\rm S}\gtrsim 10$.  Further analyses have included more processes and finite temperature effects, such as thermal masses, typically enhancing RH neutrino production and thus, going in the direction of yielding smaller values of $z_{\rm eq}$~\cite{RHNproduction}. The latest analysis, employing a closed path formalism~\cite{garbrecht}, finds for the total production rate $\G^{\rm tot}_{N_{\rm S}} (z\ll 1) \simeq 3 \times 10^{-3}\,h^2_{\rm S}\,T\,n_N^{\rm eq}(z\ll 1)$, which is equivalent to $(D+S)(z\ll 1) \simeq 0.2\,K_{\rm S}$, and implies $z_{\rm eq}\simeq 5/K_{\rm S}$.  Notice that, in principle, $z_{\rm eq}$, and consequently $z_{\rm res}$, can be made arbitrarily small by making $K_{\rm S}$ arbitrarily large, but a value $K_{\rm S} \gg m_{\rm atm}/m_{\star} \simeq 50$, corresponding to $z_{\rm eq} \ll 0.1$, necessarily involves some amount of fine-tuning in the seesaw formula. In any case, it is convenient to treat, for the time being, $z_{\rm res}$ as a free parameter since it plays a crucial role. Moreover, when we will discuss the case of dynamical $N_{\rm S}$ abundance at the resonance, we will show that it is possible to have arbitrarily small values of $z_{\rm res}$, although below $z_{\rm eq}$.

Plugging now the expression for $\g_{\rm res}$, Eq.~(\ref{gres2}), into the DM abundance, Eq.~(\ref{DMabundance}), one obtains
\be
\O_{\rm DM}\,h^2 \simeq \frac{0.14}{\a_{\rm S} \, z_{\rm res}} \, \left(\frac{M_{\rm DM}}{M_{\rm S}}\right) \,
\left(\frac{10^{20} \, {\rm GeV}}{\widetilde{\L}}\right)^2 \, \left(\frac{M_{\rm DM}}{{\rm GeV}}\right) ~.
\ee
Latest {\em Planck} satellite results find for the DM abundance (combining temperature and polarization anisotropies and gravitational lensing)~\cite{planck}, 
\be
\O_{\rm DM}\,h^2 =0.1193 \pm 0.0014  ~.
\ee
This implies that the correct value of $\widetilde{\L}$ to reproduce the observed DM abundance is given by 
\be\label{LAMBDADM}
\widetilde{\L}_{\rm DM} \simeq 10^{20} \, \sqrt{\frac{1.15}{\a_{\rm S} \, z_{\rm res}} \, \frac{M_{\rm DM}}{M_{\rm S}}\, \frac{M_{\rm DM}}{{\rm GeV}}}  \,\, {\rm GeV} ~,
\ee 
showing that the mechanism can reproduce the correct DM abundance for reasonable values of $\widetilde{\L}_{\rm DM}$.\footnote{One has values $\widetilde{\L}_{\rm DM} \equiv \L/\l_{AS} \gg 10^{20}$~GeV. If $\L_{\rm DM} \simeq M_{\rm Pl}$, quite small couplings $\l_{AS}\ll 0.1$ are required. However, even if one imposes $\l_{AS} \sim 1$, it is not difficult to build models with an effective scale of energy $\L \gg M_{\rm Pl}$. For example in GUT theories the RH neutrinos can couple, with a small Yukawa coupling $h$, to a heavy scalar $H$ with mass $M_H\sim M_{\rm GUT}$ and this via an one-dimensional trilinear coupling $\m \ll M_{\rm GUT}$ to the SM Higgs via a tadpole graph. In this case, integrating out the heavy scalar, one obtains $\L \sim M^2_{\rm GUT}/(\mu\,h) \gg M_{\rm GUT}$ \cite{unpublished}.}

In the hierarchical case, $z_{\rm res} \simeq  h_{\rm S}/2 \simeq 0.85 \times 10^{-8}\sqrt{\a_{\rm S} \, (M_{\rm DM}/{\rm GeV})}$. If $z_{\rm res}$ is set to $z_{\rm res} \gtrsim z_{\rm eq} \gtrsim 0.1$, then $M_{\rm DM} \gtrsim 10^{14}\,{\rm GeV}/\a_{\rm S}$. These values imply unacceptably high values of the reheat temperature, since it is required  $T_{\rm res}\gtrsim 10^{15}\,{\rm GeV}/\a_{\rm S}$ at the end of inflation. This was the argument used in \cite{CDMRHnumix} to rule out the hierarchical case. However, as we said, we will show that values $z_{\rm res}\ll 0.1$ are actually possible if the assumption of ultra-relativistic thermal $N_{\rm S}$ abundance at the resonance is relaxed. Interestingly, in light of current IceCube high energy neutrino data, we will see that this will rescue the possibility to open a window for $M_{\rm DM} \gg M_{\rm S}$, without any additional non-minimal ingredient or resorting to theoretical uncertainties.

On the other hand, in the quasi-degenerate case, one has $\d_{\rm DM} \equiv (M_{\rm DM}-M_{\rm S})/M_{S} \ll 1$, so that we can write
\be\label{zres2}
z_{\rm res} \simeq \frac{h_{\rm S}}{2 \, \sqrt{2\,\d_{\rm DM}}} \simeq 
6 \times 10^{-9} \, \sqrt{\frac{\a_{\rm S}}{\d_{\rm DM}} \, \left(\frac{M_{\rm DM}}{{\rm GeV}}\right)} ~.
\ee
This shows that, for a fixed value of $M_{\rm DM}$ and a sufficiently small value of $\d_{\rm DM}$, it is  always possible to satisfy the condition for ultra-relativistic thermal equilibrium, $z_{\rm res} \gtrsim 0.1$. 

We have now to worry whether the same new interactions responsible for the production of the abundance can spoil the DM stability on cosmological scales giving unacceptably short lifetimes with high-energy neutrino flux in disagreement with the IceCube data. Of course, at the same time, this instability also represents an opportunity to identify a potential observable contribution to the detected IceCube high-energy neutrinos, as first pointed out in Ref.~\cite{CDMRHnumix}. This implies that the model has predictive power and can be tested.

\subsection{DM decays}
\label{sec:DMdecays}
 
The $N^{\L,0}_{\rm DM}$ decays can proceed through two dominant decay channels~\cite{CDMRHnumix, anisimov, unpublished}. The first channel is due to the mixing itself that produces the observed DM abundance. Indeed, after electroweak spontaneous symmetry breaking, though the finite temperature-induced mixing is negligible, the operator in Eq.~(\ref{newoperator}) still generates a mixing (in vacuum) between $N_{\rm DM}$ and $N_{\rm S}$ with mixing angle\footnote{It is possible to derive this expression going through the usual lines already reviewed to derive $\theta_{\L}(T)$.}
\be\label{thetaL0}
\theta_{\L}^0 = \left(\frac{v^2}{\widetilde{\L}}\right)^2 \,
\frac{1}{\G_{\rm S}^2/4 +M^2_{\rm S}\,\d_{\rm DM}^2},
\ee
where $\G_{\rm S} \equiv h_{\rm S}^2\,M_{\rm S}/(4\,\pi)$ is the total $N_{\rm S}$ decay width for $M_{\rm DM} > M_{\rm Higgs} \simeq 125$~GeV~\cite{pascoli}.

In this way, $N^{\L,0}_{\rm DM}$ does not exactly coincide with the stable neutrino mass (and Yukawa) eigenstate $N_{\rm DM}$, but has a tiny (fast-decaying) $N_{\rm S}$ component that would decay quickly into gauge bosons and leptons. As we describe below, the flavour composition of the produced light $\nu_{\rm S}$ neutrinos could play an interesting role in the analysis of the high-energy neutrino flux predicted by the mechanism. The decay rate for this process of DM decay via mixing is then simply given by~\cite{anisimov, unpublished}
\be
\G_{{\rm DM} \ra {\rm S} \ra \phi + \nu_{\rm S}} \equiv \G_{{\rm DM} \ra \phi + \nu_{\rm S}} \simeq 
\left(\frac{v^2}{\widetilde{\L}}\right)^2 \frac{\G_S}{\G_{\rm S}^2/4 + M^2_{\rm S} \, \d_{\rm DM}^2}.
\ee
This can be translated into an expression for the DM lifetime
\be
\tau_{{\rm DM} \ra {\rm S} \ra \phi + \nu_{\rm S}} \simeq  \t_{\rm S} \, \left(\frac{\widetilde{\L}}{v}\right)^2 
\, \left(\frac{\G_{\rm S}^2}{4 \, v^2} + \frac{M_{\rm S}^2}{v^2}\,\d_{\rm DM}^2 \right)  ~,
\ee
where
\be
\t_{\rm S} \equiv \G_{\rm S}^{-1} \simeq 2.8\times 10^{-8} \, \a^{-1}_{\rm S} \, \left(\frac{{\rm GeV}}{M_{\rm S}}\right)^2 \, {\rm s} ~.
\ee
Moreover, since 
\be\label{MSsquaredDeltaDMsquared}
\left(\frac{M_{\rm S}^2}{v^2}\right) \, \d_{\rm DM}^2 = \frac{\G_{\rm S}^2}{v^2} \, 
\frac{\pi^2}{z^4_{\rm res}} \, \left(\frac{M_{\rm DM}/M_{\rm S}}{1 + M_{\rm S}/M_{\rm DM}}\right)^2
\gg \frac{\Gamma_{\rm S}}{4v^2} ~,
\ee 
the first term can be neglected and we can write, using Eqs.~(\ref{hS}) and~(\ref{DM2})
\bea\label{tau2bodybis}
\tau_{{\rm DM} \ra {\rm S} \ra A + \nu_{\rm S}} & \simeq &  \t_{\rm S} \, \left(\frac{\widetilde{\L}}{v}\right)^2 \,  \frac{M_{\rm S}^2}{v^2}\,\d_{\rm DM}^2 = 
\t_{\rm S}\, \left(\frac{\widetilde{\L}}{v}\right)^2 \,
\frac{h^4_{\rm S} \, M_{\rm DM}^2}{16\,v^2\,z_{\rm res}^4} \, \frac{1}{(1 +M_{\rm S}/M_{\rm DM})^2}  \nonumber  \\
& \simeq & 
1.6 \times 10^{-49} \, \frac{\a_{\rm S}}{z_{\rm res}^4}\,
\left(\frac{\widetilde{\L}}{{\rm GeV}}\right)^2 \,
\left(\frac{M_{\rm DM}}{{\rm GeV}}\right)^2
\, \frac{1}{(1 +M_{\rm S}/M_{\rm DM})^2} \,\,\, {\rm s} ~.
\eea
Finally, imposing the condition in Eq.~(\ref{LAMBDADM}) on $\widetilde{\L}$ in order to obtain the correct DM abundance, the DM lifetime can be written as
\be\label{tau2body}
\tau_{{\rm DM} \ra {\rm S} \ra A + \nu_{\rm S}} \simeq 
\frac{1.84 \times 10^{-9}\,{\rm s}}{z^{5}_{\rm res}} \,
\left(\frac{M_{\rm DM}}{{\rm GeV}}\right)^3  \, 
\, \frac{M_{\rm DM}/ M_{\rm S}}{(1 +M_{\rm S}/M_{\rm DM})^2}  ~.
\ee
As we discuss below, IceCube data constrain $\tau_{\rm DM} \gtrsim \tau_{\rm DM} ^{\rm min} \sim 10^{28}\,{\rm s}$ and, therefore, a lower bound on $M_{\rm DM}$ is obtained
\be\label{MDMlb}
M_{\rm DM} \geq M_{\rm DM}^{\rm min} \simeq 2.5 \times 10^{12} \, z_{\rm res}^{5/3} 
\,\t_{28}^{1/3} \,\left[\frac{(1+M_{\rm S}/M_{\rm DM})^2}{4\,M_{\rm DM}/M_{\rm S}}\right]^{1/3} \, {\rm GeV}   ~, 
\ee
where we defined $\t_{28}\equiv \t_{\rm DM}^{\rm min}/10^{28}\,{\rm s}$. There is another competing decay channel: the 4 body-decay process $N_{\rm DM} \ra 2 \, A + N_{\rm S} \ra 3\,A +\nu_{\rm S}$ ($A=W^{\pm},Z,H$).\footnote{In case of $W^\pm$ emission, $\nu_S$ would be replaced by $\ell_{\rm S}^\mp$, i.e., $N_{\rm DM} \ra 2 \, A + N_{\rm S} \ra 2\,A + W^\pm + \ell_{\rm S}^\mp$.} For $M_{\rm DM}-M_{\rm S} \gg 2\,M_{A} \simeq 200$~GeV, the decay rate for this process is approximately given by 
\be\label{rate4body}
\G_{N_{\rm DM} \ra 3\,A +\nu_{\rm S}} \simeq 
\frac{\G_{\rm S}}{15 \cdot 2^{11}\, \pi^4} \, \frac{M_{\rm DM}}{M_{\rm S}}  \left(\frac{M_{\rm DM}}{\widetilde{\L}}\right)^2 
\simeq 3.3\times 10^{-7}\,\G_{\rm S} \,  {M_{\rm DM} \over M_{\rm S}} \, \left(\frac{M_{\rm DM}}{\widetilde{\L}}\right)^2  ~,
\ee
and this implies a DM lifetime 
\be
\tau_{{\rm DM} \ra 3\,A + \nu_{\rm S}} \simeq  \frac{0.1 \,{\rm s}}{\a_{\rm S}}\,
\left(\frac{{\rm GeV}}{M_{\rm DM}}\right)^4 \, 
\left(\frac{M_{\rm DM}}{M_{\rm S}}\right) \,\left(\frac{\widetilde{\L}}{{\rm GeV}}\right)^2  ~. 
\ee
For the value of $\widetilde{\L}$ to reproduce the correct DM abundance, Eq.~(\ref{LAMBDADM}), and for $\tau_{\rm DM} \gtrsim \tau_{\rm DM}^{\rm min}$, one finds the upper bound 
\be\label{MDM+1}
M_{\rm DM} \lesssim M^{{\rm max} (A)}_{\rm DM} \simeq \frac{5 \times 10^3
\,{\rm GeV}}{\a_{\rm S}^{2/3} \, z_{\rm res}^{1/3}\,\t_{28}^{1/3}} \, \left(\frac{M_{\rm DM}}{M_{\rm S}}\right)^{2/3} ~. 
\ee
This upper bound is quite stringent but it can be  circumvented by requiring $M_{\rm DM}-M_{\rm S} \lesssim 2\,M_{A} \simeq 200$~GeV (which implies the quasi-degenerate limit $M_{\rm DM}\simeq M_{\rm S}$), since in this way the process is kinematically forbidden. This condition translates into another upper bound
\be\label{upperbound}
M_{\rm DM} \leq M^{{\rm max}(B)}_{\rm DM}  \simeq  2.5 \times 10^9 \, \frac{z_{\rm res}}{ \sqrt{\a_{\rm S}}} \, {\rm GeV} ~.
\ee
Therefore, the upper bound is given by $M_{\rm DM}^{\rm max} = {\rm max}\{M^{{\rm max}(A)}_{\rm DM},M^{{\rm max}(B)}_{\rm DM}\}$. For $z_{\rm res} \gtrsim 0.5\times 10^{-4}\,\a_{\rm S}^{-1/8}\,\t_{28}^{-1/4}$ one has $M^{{\rm max}(B)}_{\rm DM} > M^{{\rm max}(A)}_{\rm DM}$, whereas for lower values of $z_{\rm res}$ one has $M^{{\rm max}(B)}_{\rm DM} < M^{{\rm max}(A)}_{\rm DM}$ and thus, if the four-body decay channel is open, a weaker upper bound is found. Of course, in the hierarchical limit, $M_{\rm DM}\gg M_{\rm S}$, one always has $M_{\rm DM}^{\rm max}=M^{{\rm max}(A)}_{\rm DM}$.

\begin{figure}
	\begin{center}
		\includegraphics[height=0.2\textheight]{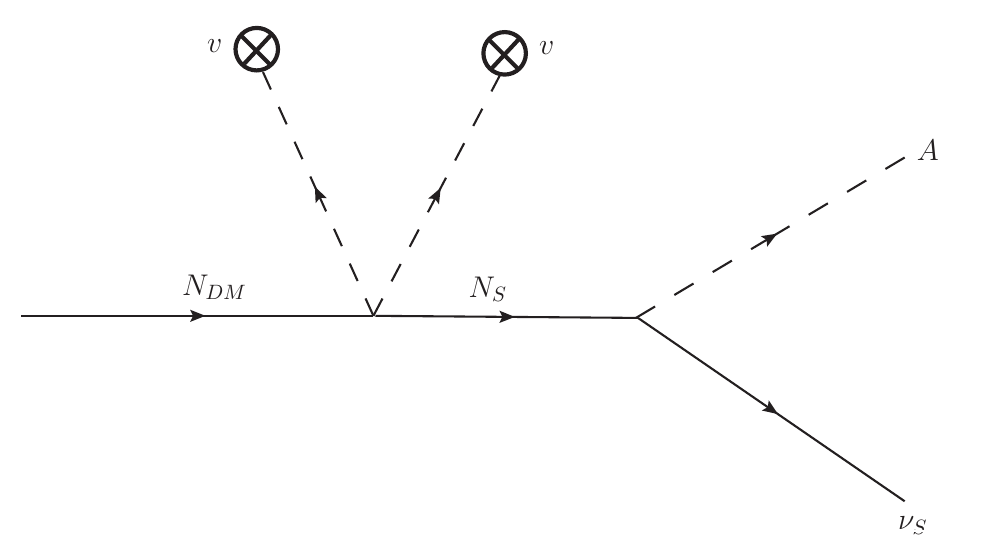}
		\includegraphics[height=0.2\textheight]{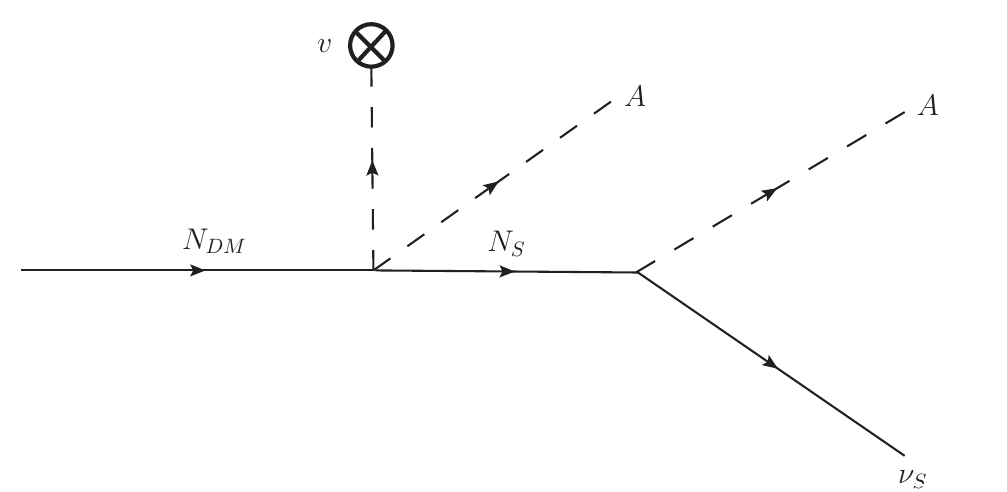}
		\includegraphics[height=0.2\textheight]{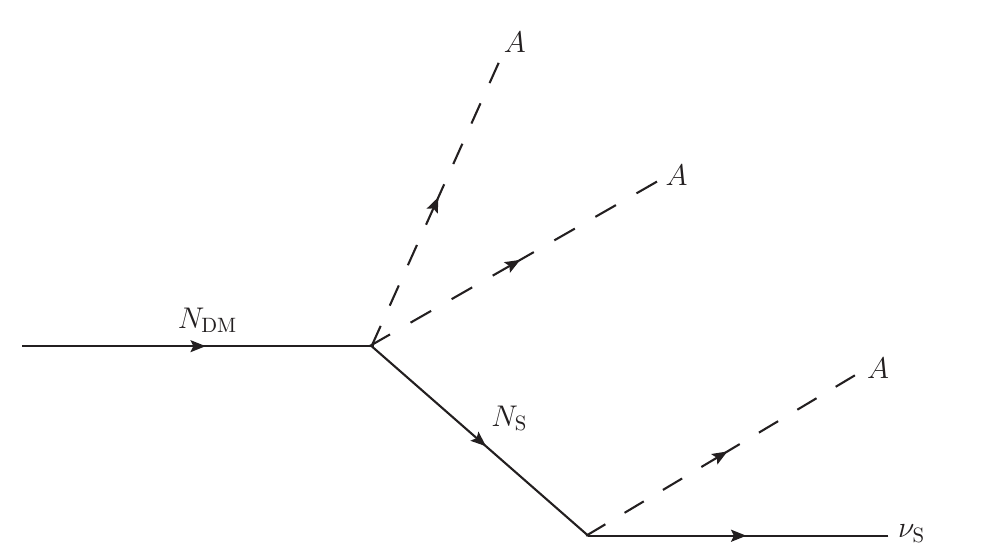}\\
	\end{center}
	\vspace{-5mm}
	\caption{Feynman diagrams of two-, three- and four-body decays of $N_{\rm DM}$.}
	\label{fig:decaychannels}
\end{figure}

Along with two- and four-body decays, there would be also three-body decays, $N_{\rm DM} \ra N_{\rm S} + A \ra 2 A + \nu_{\rm S}$. This decay channel is, however, sub-dominant with respect to four-body decays and we will not consider it. In Fig.~\ref{fig:decaychannels} we show the diagrams for the different processes in which $N_{\rm DM}$ could decay.\footnote{Annihilations $N_{\rm DM} + \bar{N}_{\rm DM} \ra 2 \, A$ should also be considered but these are subdominant, although potentially they might give a signal in dense environments in particular cases.}

The existence of solutions satisfying all discussed physical constraints necessarily requires $M_{\rm DM}^{\rm max}(\a_{\rm S}, z_{\rm res}) \geq M_{\rm DM}^{\rm min}(z_{\rm res})$ and, in this case, an allowed window, $M_{\rm DM}^{\rm min} \leq M_{\rm DM} \leq M_{\rm DM}^{\rm max}$, opens up. Let us now split the discussion, first considering the quasi-degenerate limit for $M_{\rm DM} \simeq M_{\rm S}$ and then the hierarchical limit with $M_{\rm DM} \gg M_{\rm S}$.

\subsection{The quasi-degenerate case: $M_{\rm DM}\simeq M_{\rm S}$}

Here, we describe the constraints on $M_{\rm DM}$ for the quasi-degenerate case under different assumptions, as for instance on the initial $N_{\rm S}$ abundance. As an independent parameter, in addition to $M_{\rm DM}$, in this case it is more convenient to use $z_{\rm res}$ rather than $\d_{\rm DM}$ or $M_{\rm DM}/M_{\rm S}$ or $M_{\rm S}$ and it is useful to invert Eq.~(\ref{zres2}) and write
\be
\d_{\rm DM} = \frac{h^2_{\rm S}}{8\,z^2_{\rm res}}\,  \simeq 
0.35 \times 10^{-16} \, \left(\frac{\a_{\rm S}}{z^2_{\rm res}}\right) \, \left(\frac{M_{\rm DM}}{{\rm GeV}}\right)  ~.
\ee
Let us now discuss different possibilities highlighting some issues and indicating the remedies.

\begin{itemize}
\item[(i)] {\em Initial $N_{\rm S}$ thermal abundance.}
If $N_{\rm S}$ has a thermal abundance for arbitrarily small values of $z_{\rm eq} \equiv M_{\rm DM}/T_{\rm eq}$, then $z_{\rm res}$ can be treated as independent free parameter.\footnote{As usual, such early thermalisation can be justified in terms of extra gauge interactions associated to annihilations of $Z'$ in left-right symmetric models which thermalise $N_{\rm S}$~\cite{plumacher}. Of course, in this case $N_{\rm DM}$ should be a singlet under these new gauge interactions.} In this case, an allowed window of $M_{\rm DM}$ values starts to open up when
\be\label{MDMzresfree}
M_{\rm DM}^{\rm min}  = M_{\rm DM}^{{\rm max} (A)} \equiv M_{\rm DM}^{\star}  \simeq 1.4 \times 10^5 \, \a_{\rm S}^{-{5/9}}  \,  \t_{28}^{-2/9}  \,{\rm GeV}  ~,
\ee
which is obtained for
\be
z_{\rm res} = z_{\rm res}^{\rm max} \simeq 0.45\times 10^{-4}\, 
\a_{\rm S}^{-{1/3}}  \,  \t_{28}^{-1/3}  ~,
\ee
and
\be\label{deltaDMmin(i)}
\d_{\rm DM}  = \d^{\rm min}_{\rm DM} \simeq  2 \times 10^{-3}\, \a_{\rm S}^{10/9} \,  
\t_{28}^{4/9}   ~.
\ee
Since $M_{\rm DM}^{{\rm max}(A)}\propto z_{\rm res}^{-1/3}$, for $z_{\rm res} < z_{\rm res}^{\rm max}$ the range of allowed masses enlarges and the upper bound gets more relaxed, though notice that for $z_{\rm res} \lesssim 10^{-6}$,  corresponding to $M_{\rm DM}^{\rm max} \simeq 500$~TeV, the quasi-degenerate limit does not hold any more. However, we will extend the result to the hierarchical case in Section \ref{sec:hierarchical}. In any case this scenario has the clear drawback that the required small values of $z_{\rm res}\sim 10^{-4}$  imply equally small values of $z_{\rm eq}$ and therefore, some additional interaction able to thermalise $N_{\rm S}$ (but not $M_{\rm DM}$). 

\item[(ii)] {\em Initial $N_{\rm S}$ vanishing abundance.}
Alternatively, one can wonder whether, starting from an initial vanishing $N_{\rm S}$ abundance, the $N_{\rm S}$ Yukawa interactions could produce an ultra-relativistic thermal $N_{\rm S}$ abundance, without any extra-interaction. Using $z_{\rm eq}\simeq 5/K_{\rm S} = 5\, m_{\star}/\widetilde{m}_{\rm S} = 5\,m_{\star}/(\a_{\rm S}\,m_{\rm sol})\simeq 0.5\, \a_{\rm S}^{-1}$, the larger the value of $\a_{\rm S}$, the smaller the value of $z_{\rm eq}$ and hence, the smaller the value of $z_{\rm res}$ could be (recall that $z_{\rm res}\gtrsim z_{\rm eq}$). However, the larger the value of $\a_{\rm S}$, the more stringent the upper bound $M_{\rm DM}^{\rm max}$, Eq.~(\ref{MDM+1}), and with the condition $M_{\rm DM}^{\rm min}\leq M_{\rm DM}^{{\rm max}(A)}$, an allowed range opens up when 
\be
M_{\rm DM}^{\rm min}  = M_{\rm DM}^{{\rm max}(A)} \equiv M_{\rm DM}^{\star} \simeq 100 \, \t_{28}^{-1/2} \,\, {\rm GeV} ~,
\ee
for $\a_{\rm S} \simeq 10^6 \,\t_{28}^{1/2}$. For larger values of $\a_{\rm S}$, although the allowed window for $M_{\rm DM}$ enlarges, the upper bound becomes even more stringent and our approximations for the calculation of the DM lifetime break down for different reasons. In such a case, $N_{\rm DM}$ becomes lighter than the Higgs boson and also three-body decays become dominant. Moreover, note that for such low values of $M_{\rm DM}$ sphalerons are not effective and leptogenesis is not viable. In addition, such large values of $\a_{\rm S}$ imply a huge amount of fine tuning in the seesaw formula. Therefore, this solution is not particularly appealing. 

\item[(iii)] {\em Relaxing the ultra-relativistic $N_{\rm S}$ thermal abundance condition for initial $N_{\rm S}$ vanishing abundance.} 
So far, we have assumed $N_{\rm S}$ to be in ultra-relativistic thermal equilibrium at the resonance, so that the correct abundance of $N_{\rm DM}$ is produced with the highest value of $\widetilde{\L}$ which implies the longest DM lifetime. However, the strong dependence of the lifetime\footnote{This derives from $\theta_{\L}^0 \propto z_{\rm res}^4$, Eq.~(\ref{thetaL0}) and~(\ref{MSsquaredDeltaDMsquared}), so that higher resonant temperatures imply smaller mixing between $N_{\rm DM}$ and $N_{\rm S}$.} on $z_{\rm res}$, $\t_{\rm DM \ra {\rm S} \ra \nu + \phi} \propto z_{\rm res}^{-4}$, actually suggests that if the $N_{\rm S}$ abundance is not too suppressed for  $z_{\rm res} < z_{\rm eq}$, the lower $N_{\rm S}$ abundance can be compensated by a stronger coupling (lower $\widetilde{\L}$) without spoiling the DM stability. Let us see this quantitatively. The $N_{\rm S}$ abundance can be described in terms of the kinetic equation~\cite{pedestrians}
\be
\frac{dN_{N_{\rm S}}}{dz} = - (D+S)\,(N_{N_{\rm S}}-N_{N_{\rm S}}^{\rm eq})  ~,
\ee
where $D\equiv \G_D/(H\,z)$ and $S\equiv \G^{\rm tot}_{\rm S}/(H\,z)$, with $\G_D$ and $\G^{\rm tot}_{\rm S}$  defined as the total decay and scattering rates ($2 \leftrightarrow 1$ and $2\leftrightarrow 2$ processes), respectively. For $z < z_{\rm eq} \simeq  5/K_{\rm S} \simeq 0.5\, \a_{\rm S}^{-1}$ and assuming an initial vanishing abundance, the first term describing decays can be neglected and the $N_{\rm S}$ production from the thermal bath is described simply by (using the normalisation $N_{N_{\rm S}}^{\rm eq}(z\ll 1)=1$)
\be\label{kineq}
\frac{dN_{N_{\rm S}}}{dz} =  (D+S)   ~.
\ee
Since $(D+S)(z\ll 1) \simeq K_{\rm S}/5 = 1/z_{\rm eq}$, one obtains the simple solution
\be
N_{N_{\rm S}}(z < z_{\rm eq}) \simeq \frac{z}{z_{\rm eq}}   ~,
\ee
which shows that for $z < z_{\rm eq}$ there is a linear suppression of the $N_{\rm S}$ abundance. Now, using this solution to re-write 
$(N_{N_{\rm S}}/N_{\g})_{\rm res}= 3z_{\rm res}/(4z_{\rm eq})$ in Eq.~(\ref{DMabgres}), the Eq.~(\ref{LAMBDADM}) for $\widetilde{\L}_{\rm DM}$ has to be replaced ($M_{\rm DM}\simeq M_{\rm S}$) by
\be\label{LAMBDADMdyn}
\widetilde{\L}_{\rm DM} \simeq 10^{20} \, \sqrt{\frac{1.15}{\a_{\rm S} \, z_{\rm eq}} \,
\left(\frac{M_{\rm DM}}{{\rm GeV}}\right)} \,\, {\rm GeV}  ~.
\ee
Consequently,  the lower bound, Eq.~(\ref{MDMlb}), valid for $3 \gtrsim z_{\rm res} \geq z_{\rm eq}$, becomes
\be
M^{\rm min}_{\rm DM} \simeq 2.5 \times 10^{12} \,z_{\rm eq}^{1/3}\,z_{\rm res}^{4/3} 
\, \tau_{28}^{1/3} \,\, {\rm GeV}
\hspace{5mm} (z_{\rm res}\leq z_{\rm eq})  ~,
\ee
and the upper bound, Eq.~(\ref{MDM+1}), 
\be\label{MDM+1dyn}
M^{{\rm max}(A)}_{\rm DM} \simeq 6.3 \, \a_{\rm S}^{-1/3} \, 
\t_{28}^{-1/3} \,\, {\rm TeV}  ~,
\ee
where $z_{\rm res}$ has been replaced by $z_{\rm eq}\simeq 0.5/\a_{\rm S}$. The advantage is that now $z_{\rm res}$ does not  depend on $\a_{\rm S}$ any more and can be taken arbitrarily small, albeit with the condition on the reheat temperature $T_{\rm RH} \gtrsim T_{\rm res} = M_{\rm DM}/z_{\rm res}$. Imposing again $M_{\rm DM}^{\rm min} \leq M_{\rm DM}^{{\rm max}(A)}$ and using $z_{\rm eq} \simeq 0.5/\a_{\rm S}$, this time the allowed window opens up at
\be\label{MDMstarxi1}
M_{\rm DM}^{\rm min} = M_{\rm DM}^{{\rm max}(A)} \equiv M_{\rm DM}^{\star} \simeq 6.3  \,
\a_{\rm S}^{-{1/3}}\,  \tau_{28}^{-1/3} \,\, {\rm TeV} ~,
\ee
which is realised for
\be
z_{\rm res} = z_{\rm res}^{\rm max} \simeq 4.2 \times 10^{-7}\, 
\tau_{28}^{-1/2}   ~,
\ee
and implies
\be\label{dDMminxi1}
\d_{\rm DM} \gtrsim \d_{\rm DM}^{\rm min} \simeq 1.8\,\a_{\rm S}^{2/3}\,\tau_{28}^{2/3}  ~.
\ee
This shows that this solution does not satisfy the quasi-degenerate limit, $\d_{\rm DM}\ll 1$, and has to be treated more carefully within the hierarchical case, which we discuss in the next subsection, where we see how an allowed window indeed exists and gets enlarged for $\d_{\rm DM}\gg 1$. 

If one compares this solution for $M_{\rm DM}$ with Eq.~(\ref{MDMzresfree}), obtained in the case of free $z_{\rm res}$ and ultra-relativistic thermal $N_{\rm S}$ abundance, clearly the accessible values of $M_{\rm DM}$ are more constrained, but the required small values of $z_{\rm res}$ are now perfectly justified. However, there are still a few options that have to be considered and that can rise the scale for $M_{\rm DM}$ in the quasi-degenerate case.

\item[(iv)] {\em Non-thermal $N_{\rm S}$ abundance}. 
We have so far assumed either an initial vanishing $N_{\rm S}$ abundance or a thermal abundance. One could think of a scenario in which, at the end of inflation, part of the inflaton energy density is transferred to $N_{\rm S}$'s, so they might have an initial abundance effectively much higher than their ultra-relativistic abundance \cite{giudice}. The resonant conversion could then occur in the stage when the SM content has quickly thermalised via gauge interactions, while $N_{\rm S}$'s still have a large non-thermal abundance. This possibility can be simply described by introducing a new parameter $\xi'$, such that $(N_{N_{\rm S}}/N_{\g})_{\rm res} = 3\,\xi'/4$ and $(N_{N_{\rm DM}}/N_\g)_{\rm res} = 
(N_{N_{\rm DM}}/N_{N_{\rm S}})_{\rm res}\,(3\,\xi'/4)$. In this way, the value $\widetilde{\L}_{\rm DM}$ for the energy scale for the correct DM abundance, Eq.~(\ref{LAMBDADM}), gets simply multiplied by a factor $\xi \equiv \sqrt{\xi'}$. Consequently, the lower bound $M^{\rm min}_{\rm DM}$, Eq.~(\ref{MDMlb}), is relaxed by a factor $\xi^{-2/3}$, while the upper bound $M_{\rm DM}^{{\rm max}(A)}$, Eq.~(\ref{MDM+1}), is relaxed by a factor $\xi^{2/3}$. Imposing again $M^{\rm min}_{\rm DM}(\xi) \leq M^{{\rm max}(A)}_{\rm DM}(\xi)$, now one finds an allowed window when
\be\label{MDMmaxxi}
M_{\rm DM}^{\rm min} (\xi)= M_{\rm DM}^{{\rm max}(A)} (\xi) \equiv M_{\rm DM}^{\star (A)}(\xi) \simeq 140\,  \xi^{4/9} \, \a_{\rm S}^{-{5/9}}\, \t_{28}^{{-2/9}} \,\, {\rm TeV} ~,
\ee
which is obtained for
\be\label{zresmaxxi}
 z_{\rm res} = z_{\rm res}^{{\rm max}(A)} 
 \simeq 0.45\times  10^{-5}\, \xi^{2/3} \, \a_{\rm S}^{-{1/3}} \,   
\t_{28}^{{-1/3}}  ~,
\ee
while $\d_{\rm DM}^{\rm min}$ in Eq.~(\ref{deltaDMmin(i)}) gets relaxed by $\xi^{-8/9}$.  For sufficiently large values of $\xi \gtrsim 10$, the upper bound $M_{\rm DM}^{\mathrm{max}(B)}$ is more relaxed than $M_{\rm DM}^{\mathrm{max}(A)}$ and an allowed window opens up when
\be\label{MDMmaxxiB}
M_{\rm DM}^{\rm min} (\xi)= M_{\rm DM}^{{\rm max}(B)} \equiv M_{\rm DM}^{\star (B)}(\xi) \simeq 80 \,  \xi \, \a_{\rm S}^{-{5/4}}\, \t_{28}^{{-1/2}} \,\, {\rm TeV}   ~,
\ee
for
\be
 z_{\rm res} < z_{\rm res}^{{\rm max}(B)} 
 \simeq 3\times  10^{-5} \, \xi \, \a_{\rm S}^{-{3/4}} \, \t_{28}^{{-1/2}}   ~,
\ee
with $\d_{\rm DM}^{\rm min} \simeq 3 \times 10^{-3}\,\xi^{-1}\,\a_{\rm S}^{5/4}\,\t_{28}^{1/2}$. For example, for a value $\xi \simeq 75$ (corresponding to an initial non-thermal $N_{\rm S}$ abundance $\xi'\simeq 6000$ times larger than the ultra-relativistic thermal equilibrium value
\footnote{For such a high value of initial $N_{\rm S}$ non-thermal abundance, one would have an initially $N_{\rm S}$-dominated universe, unless one has a model with a much higher number of degrees of freedom compared to the SM at very high temperatures.  
In this case the calculation of the effective potentials and all consequent
results, including how the bounds relax with $\xi$ should be revisited. Therefore, the results should be considered more robust 
for $\xi \lesssim 10$, corresponding to  $M_{\rm DM}^{\rm max} \simeq 100\,{\rm TeV}$.}) 
one has $z_{\rm res}^{\rm max}\simeq 2\times 10^{-3}$ and $M_{\rm DM}^{\rm max}\simeq 1$~PeV for $\a_{\rm S} =\t_{28} =1$. This option has the drawback that if some external mechanism generates an initial $N_{\rm S}$ abundance, the same mechanism might also directly create the final DM abundance. However, there are more (appealing) ways to justify the same results with values of the new parameter $\xi \gg 1$, which would allow accessing values of $M_{\rm DM}$ above the value $\sim 6$~TeV found in Eq.~(\ref{MDMstarxi1}) for $\xi=1$.

\item[(v)] {\em Non-standard expansion rate.} 
Another modification of the minimal scenario that can lead to an increase of the efficiency of the mechanism of DM production is given by the possibility that, at resonance, the cosmological evolution is not in the standard radiation-dominated regime with $H(T) \propto T^2$, but expansion is slower and $H_{\rm res}$ is smaller, which would imply the resonant conversion to be more adiabatic (cf. Eq.~(\ref{gres0})). For example, this could happen during a phase transition. In this case, however, entropy production could dilute the $N_{\rm DM}$ abundance. Therefore, the initially produced DM abundance could be larger by a factor $\xi'$, corresponding to an increase of $\widetilde{\L}_{\rm DM}$ by a factor $\xi=\sqrt{\xi'}$, and thus, a longer DM lifetime (weaker couplings) is possible. Numerically, the conclusions are the same as those in point (iv), simply with a different physical interpretation for $\xi$. Of course, such a non-standard expansion rate is not easy to motivate and should be understood as a caveat we mentioned for completeness.

\item[(vi)] {\em Theoretical uncertainties: improved kinetic description might enhance the efficiency of the mechanism.}
Our description of non-adiabatic transitions, $N_{\rm S} \ra N_{\rm DM}$, is based on the Landau-Zener formula in the monochromatic approximation at zero temperature. These results should be checked within a more rigorous quantum kinetic formalism, which accounts for different effects such as finite temperature effects, that might reasonably go into the direction of enhancing the fraction of $N_{\rm S}$ converted into $N_{\rm DM}$. From this point of view, it should be noticed from Eqs.~(\ref{NDM2NS}), (\ref{gres2}) and (\ref{LAMBDADM}), that, at the resonance, one has typical values $(N_{\rm DM}/N_{\rm S})_{\rm res} \sim 10^{-7} \, ({\rm GeV}/M_{\rm DM})$, i.e. just a tiny fraction of $N_{\rm S}$'s is converted into $N_{\rm DM}$, so one could legitimately wonder whether subtle effects might actually play an important role in the calculation of the correct $N_{\rm DM}$ abundance. In this case, a value of the parameter $\xi \neq 1$, introduced in (iv), can also be regarded as a parameterisation of the theoretical uncertainties and values $\xi \lesssim 10$ cannot be excluded.

\item[(vii)] {\em Running of $\widetilde{\L}$ at low energies.} 
Because of radiative corrections, arising within a specific model from the presence of possible states between the high resonance energy scale
and low energies, the value of $\widetilde{\L}$ might vary from very high temperatures at resonance to low energies, and in particular it can increase leading to longer DM lifetimes. This effect would reconcile the DM abundance and DM stability conditions at higher values of $M_{\rm DM}$ and it can again be encoded in terms of the parameter $\xi$ introduced in point (iv) so that the same numerical arguments apply. This possible physical justification of values $\xi \gg 1$ seems to us quite plausible.

\item[(viii)] {\em $N_{\rm DM}$ as a sub-dominant DM component.} 
If $N_{\rm DM}$ constitutes only a fraction $\xi'' < 1$ of the observed DM abundance, i.e. $\O_{N_{\rm DM}} = \xi''\,\O_{\rm DM}$, and since the neutrino flux from $N_{\rm DM}$ decays is proportional to $\O_{N_{\rm DM}}/\tau_{\rm DM}$, the lower limit on $\tau_{\rm DM}$ is correspondingly reduced by a factor $\xi''$. Thus, we can identify $(\xi'')^{-2}$ with the parameter $\xi$ in point (iv). In this case, although the mechanism is not able to reproduce the whole DM relic density, it would still be motivated by the possibility to reproduce the matter-antimatter asymmetry via leptogenesis with testable signatures in neutrino telescopes.

\item[(ix)] {\em Initial vanishing $N_{\rm S}$ abundance and $\xi \neq 1$.} 
As we have seen in point (iii), for $z_{\rm res}<z_{\rm eq}$ and initial vanishing abundance (lower dynamical $N_{\rm S}$ abundance at resonance), the produced DM abundance, though suppressed, can still reproduce the observed one. In this case, small values of $z_{\rm res}$ are perfectly justified. We now consider the calculation in point (iii), but introducing the parameter $\xi$, that encodes different possible physical effects. For $\xi > 1$, one obtains
\be\label{LAMBDADMdynxi}
\widetilde{\L}_{\rm DM}(\xi) \simeq \xi\, 10^{20} \, \sqrt{\frac{1.15}{\a_{\rm S} \, z_{\rm eq}}\, 
\left(\frac{M_{\rm DM}}{{\rm GeV}}\right)} \,\, {\rm GeV}  ~,
\ee
and thus,
\be
M^{\rm min}_{\rm DM}(\xi) \simeq 2.5 \times 10^{12} \,z_{\rm eq}^{1/3}\,z_{\rm res}^{4/3} 
\, \tau_{28}^{1/3} \, \xi^{-{2/3}} \,\, {\rm GeV}
\hspace{5mm} (z_{\rm res}\leq z_{\rm eq})  ~,
\ee
\be\label{MDM+1dynxi}
M^{{\rm max}(A)}_{\rm DM} (\xi)\simeq 6.3  \, \xi^{2/3}\,\a_{\rm S}^{-{1/3}} \, 
\tau_{28}^{-1/3} \, {\rm TeV}  ~,
\ee
while $M_{\rm DM}^{{\rm max}(B)}$ does not change and for realistic values, $\xi \lesssim 100$, $M^{\mathrm{max}(A)}_{\rm DM} < M_{\rm DM}^{{\rm max}(B)}$. In this way, imposing as usual $M^{\rm min}_{\rm DM} \leq M_{\rm DM}^{{\rm max}(A)}$, an allowed window opens up at 
\be
M^{\rm min}_{\rm DM}(\xi)=M^{{\rm max}(A)}_{\rm DM}(\xi)=M^{\star}_{\rm DM}(\xi) \simeq 
6.3 \, \xi^{2/3} \, \a_{\rm S}^{-{1/3}}\,  \tau_{28}^{-1/3} \,\, {\rm TeV} ~,
\ee
which is obtained for
\be
z_{\rm res} \leq z_{\rm res}^{\rm max} \simeq 4.2 \times 10^{-7} \, \tau_{28}^{-1/2} \xi   ~,
\ee
and $\d_{\rm DM}^{\rm min}$ gets relaxed by a factor $\xi^{-4/3}$ with respect to the value found for $\xi=1$, Eq.~(\ref{dDMminxi1}). For moderate values $\xi \lesssim 15$, one has $M^\star_{\rm DM} \lesssim 40$~TeV and $\d_{\rm DM} \lesssim 0.05$, so that one is correctly in the assumed quasi-degenerate limit. We will see however that, generalising the discussion to the hierarchical case, for $\d_{\rm DM}\gg 1$, much higher values of $M_{\rm DM}$ can be easily reached even for $\xi = 1$.  

\item[(x)] {\em Mixing with a second thermalised RH neutrino $N_{\rm I}$?} 
We have assumed that the new interactions, thanks to the coupling $\l_{AS}$, mix the mass eigenstate $N_{\rm DM}$ only with $N_{\rm S}$, but not with $N_{\rm I}$, which implies a negligible coupling $\l_{AI}$. One might wonder whether turning on such a coupling might have a beneficial effect, somehow helping to relax the tension. However, at this stage, it should be clear that things can only get worse with a second coupling, since all the same constraints would also apply to this second mixing with an effective scale $\widetilde{\L}_{\rm I} \equiv \L/\l_{A {\rm I}}$. Since all constraints coming from the mixing with $N_{\rm S}$ are weakened by taking the minimal value of $\widetilde{m}_{\rm S}$, it is easy to see that it would actually be impossible to have simultaneously a minimal $\widetilde{m}_{\rm I}$. Therefore, necessarily we have to assume $\l_{A {\rm I}}\ll \l_{A \rm S}$ (or equivalently $\widetilde{\L}_{\rm I} \gg \widetilde{\L}$), so that the second mixing is negligible.   

\end{itemize}

We have seen that, assuming the quasi-degenerate limit, $M_{\rm S} \simeq M_{\rm DM}$, the requirement of simultaneously reproducing the correct DM relic abundance and satisfying the stability conditions is possible, either for initial thermal  $N_{\rm S}$ abundance (with $M_{\rm DM} \lesssim 500$~TeV for $\xi=1$) or for initial vanishing $N_{\rm S}$ abundance but only if $\xi \gg 1 $ (in this case $M_{\rm DM}\lesssim 6 \, \xi^{2/3}$~TeV). On the other hand, as we going to show in Section.~\ref{sec:hierarchical}, even for $\xi = 1$, when relaxing the quasi-degeneracy assumption, much higher values of $M_{\rm DM}$ are possible.

\subsection{The hierarchical case ($M_{\rm DM}\gtrsim 2\, M_{\rm S}$)} 
\label{sec:hierarchical}

If we let $M_{\rm DM} \gtrsim 2\, M_{\rm S}$, the bounds obtained in the quasi-degenerate case get considerably relaxed.  Indeed, assuming ultra-relativistic thermal equilibrium at the resonance, one can see from Eq.~(\ref{MDMlb}) that $M^{\rm min}_{\rm DM} \propto (M_{\rm S}/M_{\rm DM})^{1/3}$ and from Eq.~(\ref{MDM+1}) that $M^{{\rm max}(A)}_{\rm DM} \propto (M_{\rm DM}/M_{\rm S})^{2/3}$, so both the lower and the upper bound get relaxed.  This time, it is more convenient to use $M_{\rm DM}/M_{\rm S}$ (in addition to $M_{\rm DM}$) as independent parameter, rather than $z_{\rm res}$, which are related via Eq.~(\ref{zres}). Analogously to the quasi-degenerate case, now we determine the critical values for $M_{\rm DM}$ for which an allowed window opens up, for the different cases already discussed in the quasi-degenerate case, including those that can be parameterised in terms of $\xi \neq 1$.

\begin{figure}
	\begin{center}
		\includegraphics[width=0.53\textwidth]{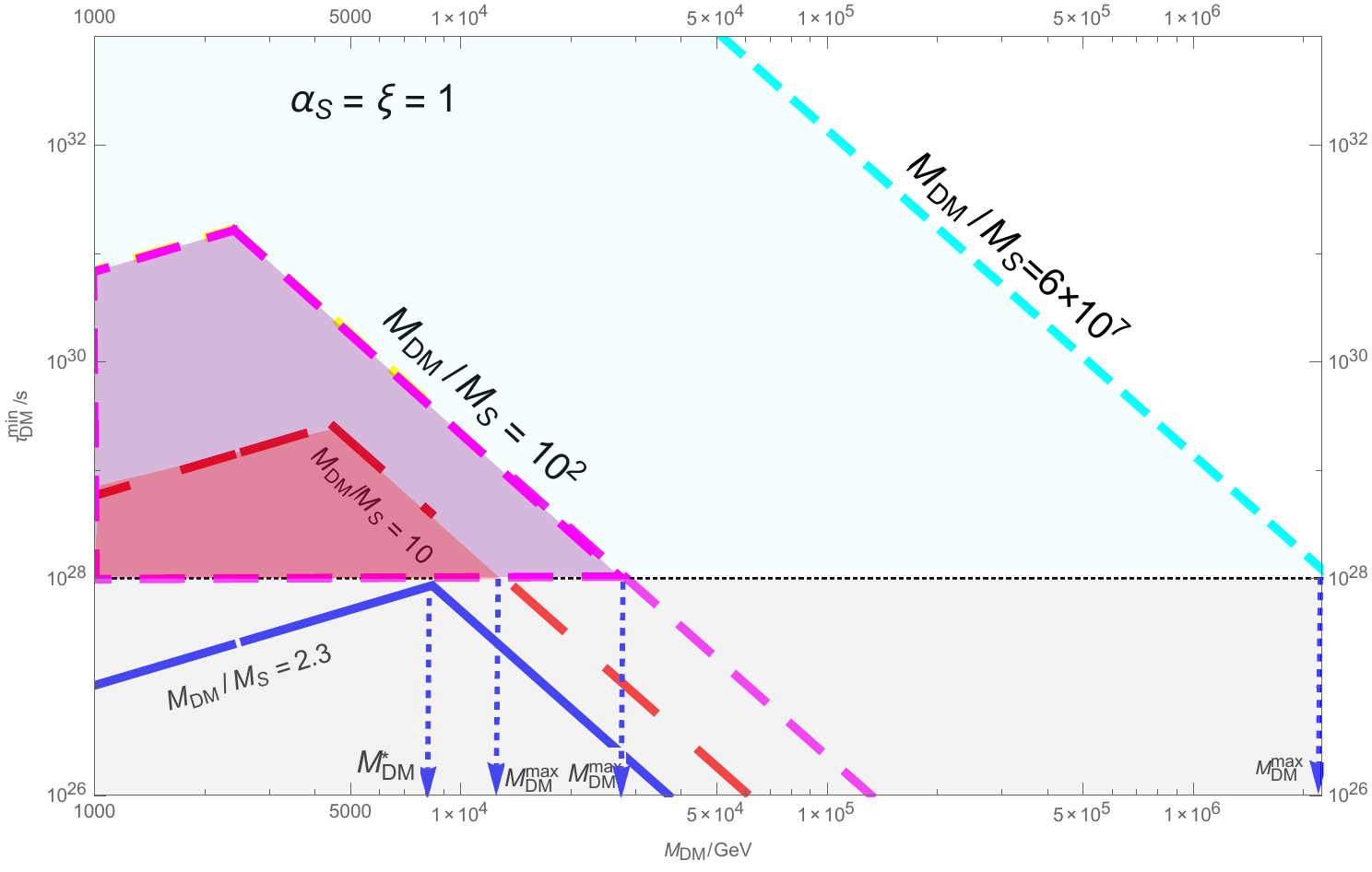} \\
		\vspace{-1mm}
		\includegraphics[width=0.53\textwidth]{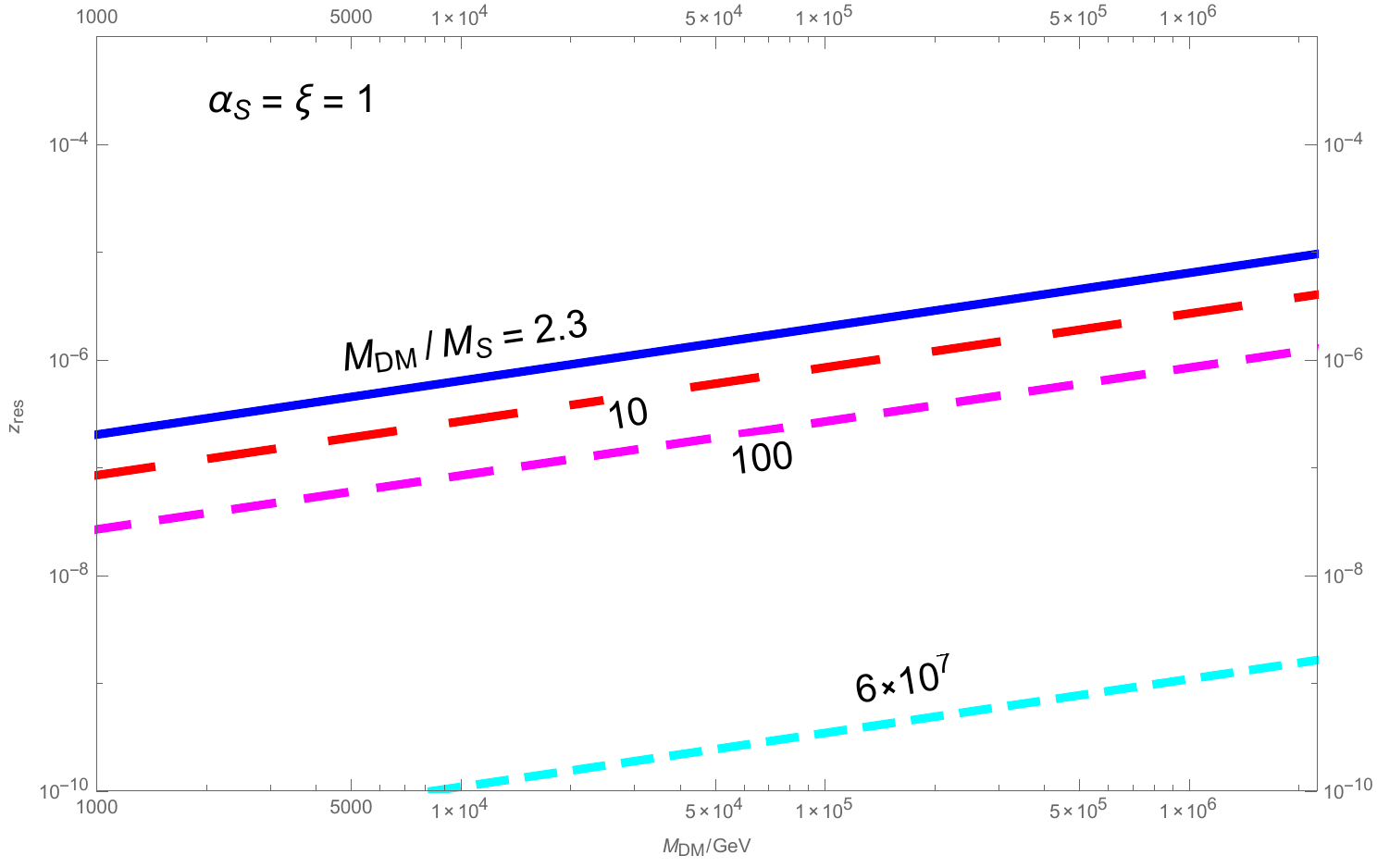} \\
		\vspace{-1mm}
		\includegraphics[width=0.53\textwidth]{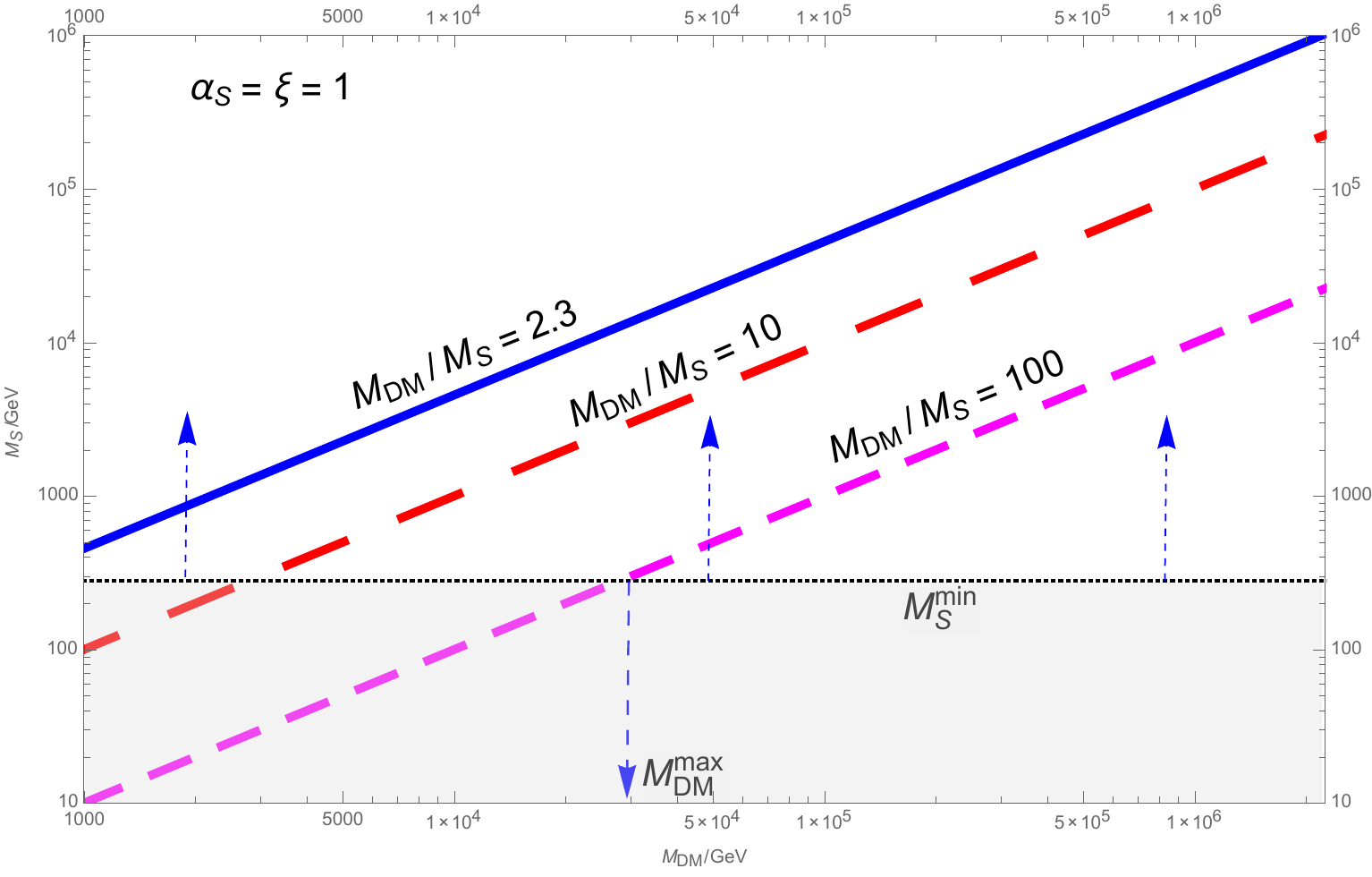} 
	\end{center}
	\vspace{-9mm}
	\caption{\small {\bf Top panel}: Bounds on $M_{\rm DM}$ in the plane $(M_{\rm DM}, \tau_{\rm DM}^{\rm min})$ for $M_{\rm DM}/M_{\rm S}$= 2.3 (blue line), 10 (red line), $100$ (magenta line) and $6 \times 10^7$ (cyan line), as indicated.  The tiny dashed horizontal line is the reference value $\tau_{\rm DM}^{\rm min}=10^{28}\,{\rm s}$. The allowed regions (in the same respective light colours) 	 are then obtained as a combination of the lower bound $M_{\rm DM}^{\rm min}$ from two body decay Eq.~(\ref{lbdynh}), visible for $M_{\rm DM}/M_{\rm S} = 2.2, 10$ and $100$,  with the upper bound from four-body decay Eq.~(\ref{ubMS}) for $\tau_{\rm DM}^{\rm min}=10^{28}\,{\rm s}$. {\bf Central panel}: $z_{\rm res}$ as a function of $M_{\rm DM}$ from Eq.~(\ref{zres}) for the same values of $M_{\rm DM}/M_{\rm S}$ as in the top panel.  {\bf Bottom panel}: $M_{\rm S}$ as a function of $M_{\rm DM}$ for the same values of $M_{\rm DM}/M_{\rm S}$ as in the top panel. The grey area is the region excluded by the lower bound $M_{\rm S} \gtrsim 2\,T_{\rm sph}^{\rm out} \simeq 300\,{\rm GeV}$ from leptogenesis.}
	\label{fig:bounds}
\end{figure}

Let us first consider the case of {\em initial   thermal $N_{\rm S}$ abundance}. One finds a trivial lower bound $M^{\rm min}_{\rm DM} \ll {\rm GeV}$, while there is still an upper bound
\be\label{ubthermal}
M_{\rm DM}^{{\rm max}}  \simeq 330 \,\a_{\rm S}^{-5/7}\,\t_{28}^{-2/7}\,\xi^{4/7}  
\left({{M_{\rm DM}\over M_{\rm S}}}\right)^{3/7}
\,\left[\left({M_{\rm DM}\over M_{\rm S}}\right)^2-1\right]^{1/7} \,\, {\rm TeV} ~,
\ee
that relaxes $ \propto (M_{\rm DM}/M_{\rm S})^{5/7}$ for increasing $M_{\rm DM}/M_{\rm S}$. On the other hand, from Eq.~(\ref{zres}), one can see that $z_{\rm res} \propto (M_{\rm DM}/M_{\rm S})^{-1/7}$, so when imposing $T_{\rm res} \lesssim 10^{15}$~GeV, the upper bound $M_{\rm DM}/M_{\rm S} \lesssim 7 \times 10^4$ is obtained, which implies $M_{\rm DM}^{\rm max}\lesssim 10^9\,{\rm GeV}\,\xi^{1/3}$. However, as already mentioned for the quasi-degenerate limit, assuming an initial thermal $N_{\rm S}$ abundance is not a particularly attractive case. 

Let us now consider the more interesting case of {\em initial vanishing $N_{\rm S}$ abundance} already considered
in the quasi-degenerate case (see Section 2.4, case (iii)).  This time one has to be careful  in noticing that the kinetic equation (\ref{kineq})
is still valid but with the replacement $z \rightarrow z_{\rm S} \equiv M_{\rm S}/T = z \, M_{\rm S} / M_{\rm DM}$.  For this reason this time one has  that the $N_{\rm S}$ abundance at the resonance is given by 
\be\label{NSzreshier}
N_{\rm S}(z_{\rm res}) = 
{z_{\rm res}\over z_{\rm eq}}\,{M_{\rm S}\over M_{\rm DM}} \,  .
\ee 
Taking into account this effect and taking into account
one (or more) of the possible effects discussed in the quasi-degenerate case and that can be all together 
parameterised in terms of a coefficient $\xi$, one obtains
\be
\widetilde{\L}_{\rm DM} \simeq 10^{20}\,\xi \, \sqrt{\frac{1.15}{\a_{\rm S} \, z_{\rm eq}} \, \left(\frac{M_{\rm DM}}{{\rm GeV}}\right)} \,\, {\rm GeV} ~.
\ee 
From this equation, using the Eq.~(\ref{zres}) to express $z_{\rm res}$ in terms of $M_{\rm DM}/M_{\rm S}$, one finds the lower bound
\be\label{lbdynh}
M_{\rm DM}^{\rm min} \simeq 14 \, \xi^{-2}\, \a_{\rm S}\,\t_{28} \,\left(\frac{M_{\rm DM}}{M_{\rm S}}-1 \right)^{ -2}  \,\, {\rm TeV} ~. 
\ee
If we now consider the upper bound from four-body decays, the upper bound found in point (ix) for the quasi-degenerate case gets relaxed by a factor $(M_{\rm DM}/M_{\rm S})^{1/3}$. Explicitly,
\be\label{ubMS}
M_{\rm DM}^{\rm max} \simeq 6.3 \, \xi^{2/3}\,\a_{\rm S}^{-1/3}\,\t_{28}^{-1/3} 
\, \left(\frac{M_{\rm DM}}{M_{\rm S}}\right)^{1/3}  \,\, {\rm TeV} ~.
\ee
If we again impose $T_{\rm res} \lesssim 10^{15}\,{\rm GeV}$, combining Eq.~(\ref{zres}) with (\ref{ubMS}), we obtain this time the upper bound $M_{\rm DM}\lesssim 2 \,{\rm PeV}\,\xi^{1/2}\,\t_{28}^{-1/4} $ corresponding to
$M_{\rm DM}/M_{\rm S} \lesssim 6 \times 10^7\,\xi^{-1/2}\,\t_{28}^{1/4}\,\a_{\rm S}$.

The situation (for initial vanishing abundance) is summarised in Fig.~\ref{fig:bounds} for $\a_{\rm S}=\xi=1$. The different constraints in the plane $\tau_{\rm DM}$ vs. $M_{\rm DM}$ are shown in the top panel for different values of $M_{\rm DM}/M_{\rm S}$. We show the allowed regions (in color) for a conservative value $\tau_{28}=1$.

First of all one can see that a window starts to open up for $M_{\rm DM}/M_{\rm S} \gtrsim 2.2$ at $M_{\rm DM}^{\star} \simeq 8\,{\rm TeV}$. In the central panel we plotted $z_{\rm res}$ vs. $M_{\rm DM}$ for the same values of $M_{\rm DM}/M_{\rm S}$. From this plot, one can easily determine $T_{\rm res}$. The figure also confirms the upper bound $M_{\rm DM}/M_{\rm S} \lesssim 6 \times 10^7$, which corresponds to $M_{\rm DM}^{\rm max} \lesssim 2 \,{\rm PeV}$ from the requirement $T_{\rm res} \lesssim 10^{15}\,{\rm GeV}$.

In conclusion, we can say that the scenario of DM from RH neutrino mixing, implies a natural window on $M_{\rm DM}$ that is quite an interesting feature of the model, since it naturally predicts high-energy neutrinos from DM decays in the energy range explored by IceCube, as first noticed in Ref.~\cite{CDMRHnumix}. Intriguingly, as we are going to discuss, this also links neutrinos at the high energies detected by IceCube to $\sim$~TeV leptogenesis.

\section{Matter-antimatter asymmetry from leptogenesis}
\label{sec:asymmetry}

So far, for the $N_{\rm DM}$ production, we have considered only two mixed RH neutrinos with a mass splitting $\delta_{\rm DM}$.  Now, we also want to take into account the presence of the third interfering RH neutrino, $N_{\rm I}$, with mass $M_{\rm I}$ (in any case necessary to reproduce correctly the solar and atmospheric neutrino mass scales), in order to have an interference with the source RH neutrino $N_{\rm S}$ giving rise to non-vanishing $C\!P$ asymmetries for the generation of a matter-antimatter asymmetry via leptogenesis \cite{CDMRHnumix}. As we explained in point (x) in the previous section, it is better to have negligible mixing between $N_{\rm DM}$ and $N_{\rm I}$ in order not to increase DM instability.

Since the matter-antimatter asymmetry of the universe is observed today in the form of a baryon asymmetry, this is related to the baryon abundance
measured by the {\em Planck} satellite given by \footnote{This is the result obtained by combining {\em Planck} satellite data on temperature and polarization anisotropies and lensing.} $\O_{\rm B,0}\,h^2 = 0.02226 \pm 0.00016$~\cite{planck}. This can be simply converted into the baryon-to-photon number ratio using
\be
\eta_{B,0} = \frac{\rho_{\rm c,0}\,h^{-2}}{m_{\rm N} \, n_{\g,0}}\simeq 273.3 \times 10^{-10}\,\O_{ B,0}\,h^2  = (6.08 \pm 0.04) \times 10^{-10} \,   ,
\ee
where $\rho_{\rm c,0}= (1.05375 \pm 0.00013) \times 10^{-5} \, h^2$~GeV/cm$^3$ is the critical energy density of the universe at the present time, $h = 0.6751 \pm 0.0064$~\cite{planck} is the Hubble constant, $H_0$, in units of $100 \, {\rm km} \, {\rm s}^{-1} \, {\rm Mpc}^{-1}$ and $m_{\rm N}$ is the nucleon mass. Let us see how we can explain this number with leptogenesis at the energy scale enforced by the DM constraints discussed in the previous section. 
 
Since the mixing angle $\theta_\L(T) $ is tiny, now we can completely neglect the mixing due to the non-standard interactions responsible for $N_{\rm DM}$ production and focus just on the interference between $N_{\rm S}$ and $N_{\rm I}$. Moreover, since $N_{\rm DM}$ has to be heavier than $N_{\rm S}$, in principle, there are two cases: either $M_{\rm DM} = M_2$ or $M_{\rm DM}= M_3$.  In the first case, leptogenesis would occur through the interference between the lightest and the heaviest RH neutrino. In the second case, it would occur via the interference of the two lightest RH neutrinos. The model is effectively a two-RH neutrino model, since $N_{\rm DM}$ is basically decoupled. 

A detailed analysis of leptogenesis in the hierarchical case of the two-RH neutrino model~\cite{2RHN} results in a lower bound, $M_1 \gtrsim 3\times 10^{10}$~GeV in the NH case and $M_1 \gtrsim 10^{11}$~GeV in the IH case, and thus, the well known lower bound~\cite{di} is in this case even more stringent. Therefore, since from the DM abundance analysis $M_{\rm DM} \neq M_1$ and $M_{\rm S}$ needs to be well below $10^{10}$~GeV, at most at the PeV scale, then one is necessarily lead to consider the quasi-degenerate limit for $N_{\rm S}$ and $N_{\rm I}$, so that the RH neutrino $C\!P$ asymmetries are enhanced~\cite{crv, pilaftsis}. 
  
\begin{figure}
\begin{center}
\psfig{file=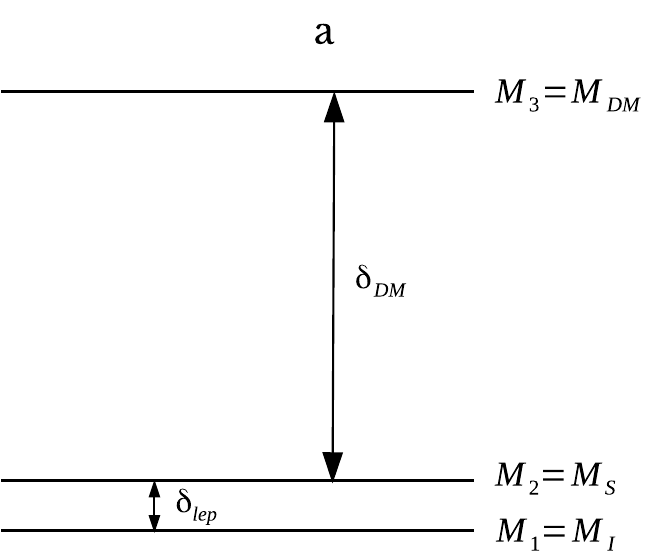,height=45mm} 
\hspace{15mm}
\psfig{file=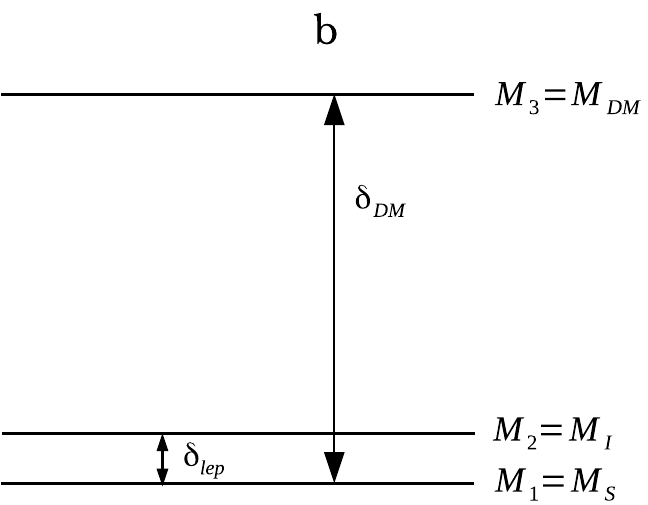,height=45mm}
\end{center}
\vspace{-5mm}
\caption{Possible RH neutrino mass spectra.}
\label{fig:spectrum}
\end{figure}

Results on leptogenesis beyond the hierarchical limit, taking into account flavour effects and assuming a two RH neutrino model, were presented in Ref.~\cite{bounds}. It was shown that, in the degenerate limit,  when $\d_{\rm lep} \lesssim 10^{-2}$~\cite{beyond}, the lower bound on $M_1$ is very similar to that one for the unflavoured case, just a factor two weaker, and the asymmetry is $\propto 1/\d_{\rm lep}$. Therefore, in our case, since the lower bound has to be relaxed by about five orders of magnitude, we can anticipate that $\d_{\rm lep} \lesssim 10^{-5}$. As we discussed, in order to satisfy all DM constraints $\d_{\rm DM} \gtrsim 10^{-2} $ for initial vanishing $N_{\rm S}$ abundance
with $\xi \lesssim 100$ and thus, necessarily  $N_{\rm DM}=N_3$, as $\d_{\rm lep}\ll \d_{\rm DM}$. Of course, there are still two possibilities: either $N_{\rm S}$ is the lightest state, i.e., $M_{\rm S} = M_1$ and $M_{\rm I} = M_2$, or the next-to-lightest, i.e., $M_{\rm S} = M_2$ and $M_{\rm I}=M_1$. These two possible cases for the RH neutrino mass spectrum are shown in Fig.~\ref{fig:spectrum}.

Therefore, the interference between the two lightest RH neutrinos generates the matter-antimatter asymmetry. The Dirac neutrino mass matrix and, correspondingly, the orthogonal matrix, are then given by the third case in Eq.~(\ref{WASS}) and Eq.~(\ref{orthogonal}), respectively. 

Let us verify this estimate by performing a quantitative analysis. Since, as we have seen, the upper bound $M_{\rm DM}^{\rm max}$ requires $M_{\rm S} \lesssim 1$~PeV, even allowing for a large $\xi$, then leptogenesis necessarily occurs in the fully three-flavoured regime~\cite{flavour}, so that the asymmetry is the sum of the three contributions from the three charged lepton flavours. At the same time, the asymmetry is the sum of the contribution from the lightest RH neutrino, $N_1$, and the contribution from the next-to-lightest RH neutrino, $N_2$. Therefore, we can write the final asymmetry as
\be\label{NBmLf}
N_{B-L}^{\rm f} = \sum_{\a}\, (N_{\D_\a}^{(1)} + N_{\D_\a}^{(2)}) \,  \hspace{5mm} (\a=e,\mu,\t)  ~.
\ee
The six individual different contributions can be expressed as 
\be
N_{\D_\a}^{(i)} = \ve_{i\a}\,\k(K_{1\a}+K_{2\a})  ~,
\ee
where we have introduced the flavoured decay parameters
\be
K_{i\a}\equiv \frac{\G_{i\a} + \overline{\G}_{i\alpha}}{H(T=M_i)} = \frac{|m_{D\a i}|^2}{M_i \, m_{\star}} ~,
\ee
and $\G_{i\a}$ and $\overline{\G}_{i\alpha}$ are the flavoured decay rates into leptons and anti-leptons, respectively. The equilibrium neutrino mass is given by
\be
m_{\star} \equiv \frac{16\,\pi^{5/2}\,\sqrt{g_{\star}^{\rm SM}}}{3\,\sqrt{5}} \, \frac{v^2}{M_{\rm Pl}} \simeq 1.1 \,\, {\rm meV} ~.
\ee
The efficiency factors $\k(K_{1\a}+K_{2\a})$ can be calculated using
\be\label{kappa}
\k(x) = \frac{2}{z_B(x)\,x}
\left(1-e^{-\frac{x\,z_B(x)}{2}}\right) \;\;\;\;\;\;
\mbox{\rm with} \;\;\;\;\;\;
z_B(x) \simeq  2+4\,x^{0.13}\,e^{-\frac{2.5}{x}}  ~,
\ee
where, in our case, $x=K_{1\a}+K_{2\a}$. This simple expression is strictly valid for initial RH neutrino thermal abundance but since, in any case, the wash-out for the two RH neutrinos adds up and it is necessarily strong in each flavour, there is basically no dependence on the initial RH neutrino abundance. Indeed, notice that, since $\d_{\rm lep} \lesssim 0.01$, we are in the degenerate limit, where the wash-out of the two RH neutrinos adds up~\cite{beyond}. The flavoured $C\!P$ asymmetries are defined as ($i,j=1,2$ and $i\neq j$)
\be
\ve_{i\a}\equiv - \frac{\G_{i\alpha}-\overline{\G}_{i\alpha}}{\G_{i}+\overline{\G}_{i}}  ~,
\ee
where $\G_{i}+\overline{\G}_{i} = \sum_{\a}(\G_{i\a} + \overline{\G}_{i\alpha}) $ are the total decay rates and the decay parameters are defined as $K_i \equiv (\G_{i}+\overline{\G}_{i})/H(T=M_i) = \sum_\a K_{i\a}$. They can be calculated using~\cite{crv}
\be\label{epsia}
\ve_{i\a} \simeq \frac{\overline{\ve}(M_i)}{K_i}
\, \left\{ {{\cal I}_{ij}^{\a}}\,\x({M^2_j/ M^2_i}) + 
{{\cal J}_{ij}^{\a}} \, \frac{2}{3(1-M^2_i/M^2_j)}\right\}  ~,
\ee
where we introduced
\be
\overline{\ve}(M_i) \equiv \frac{3}{16\,\pi} \, \left(\frac{M_i\,m_{\rm atm}}{v^2}\right) 
\simeq 1.0 \times 10^{-6} \, \left(\frac{M_i}{10^{10}\,{\rm GeV}}\right)  ~, 
\ee
\be
\xi(x)=\frac{2}{3}x\left[(1+x)\ln\left(\frac{1+x}{x}\right)-\frac{2-x}{1-x}\right]  ~,
\ee
and
\be
{\cal I}_{ij}^{\a} \equiv  \frac{{\rm Im}\left[m_{D\a i}^{\star} \, m_{D\a j} \, (m_D^{\dag} \, m_D)_{i j}\right]}{M_i \, M_j \, m_{\rm atm} \, m_{\star} } ~,
\hspace{1cm}
{\cal J}_{ij}^{\a} \equiv  
\frac{{\rm Im}\left[m_{D\a i}^{\star} \, m_{D\a j} \, (m_D^{\dag} \, m_D)_{j i}\right]}{M_i \, M_j \, m_{\rm atm} \, m_{\star} } \, \frac{M_i}{M_j}   ~.
\ee

Since, in the degenerate limit, the efficiency factor is the same for both RH neutrino contributions (in each flavour), we can rewrite Eq.~(\ref{NBmLf}) as
\be\label{NBmLf2}
N_{B-L}^{\rm f} = \sum_\a \, (\ve_{1\a}+\ve_{2\a})\,\k(K_{1\a}+K_{2\a}) \,   .
\ee
Moreover, considering that $ {\cal I}_{ij}^{\a}=-{\cal I}_{ji}^{\a}$ and that in the degenerate limit ${\cal J}_{ij}^{\a}\simeq -{\cal J}_{ji}^{\a}$ and  $\xi(M^2_i/M^2_j) \simeq (3\,\d_{\rm lep})^{-1} \simeq - \xi(M^2_j/M^2_i)$, the two RH neutrino contributions, for each flavour $\a$, add up (they do not cancel out) and
\be
N_{B-L}^{\rm f} \simeq \frac{\overline{\ve}(M_1)}{3\,\d_{\rm lep}}
\left(\frac{1}{K_1} + \frac{1}{K_2}\right)\,
\sum_{\a}\,\k(K_{1\a}+K_{2\a})\,\left[{\cal I}_{12}^{\a}+ {\cal J}_{12}^{\a}\right]  ~.
\ee

We can now write the different quantities using the orthogonal parameterisation, since this allows us to specify clearly the dependence on the low-energy neutrino parameters. The total and flavoured decay parameters can be written as
\be
K_i = \sum_j \, \frac{m_j}{m_{\star}}\, |\O_{ji}|^2  \hspace{5mm} \mbox{\rm and} \hspace{5mm}
K_{i\a} = \left|\sum_j\,\sqrt{\frac{m_j}{m_{\star}}} \, U_{\a j} \, \O_{j i}\right|^2  ~.
\ee
We can also write
\bea
{\cal I}_{ij}^{\a} & = &   \sum_{k,l,m} \, \frac{m_k\,\sqrt{m_l\, m_m}}{m_{\rm atm}\,m_{\star} }\,
\mathrm{Im}\left[
U^{\star}_{\a m}\,U_{\a l} \, \O^{\star}_{m i}\,\O_{l j}\, \O^{\star}_{k i}\,\O_{k j}\right]  ~, \\
{\cal J}_{ij}^{\a} & = & \, \sum_{k,l,m} \, \frac{m_k\,\sqrt{m_l\, m_m}}{m_{\rm atm}\,m_{\star}}\,
\mathrm{Im}\left[
U^{\star}_{\a m}\,U_{\a l}\,
\O^{\star}_{m i}\,\O_{l j}\, \O^{\star}_{k j}\,\O_{k i}\right] ~. 
\eea
Finally, defining
\be
f(m_{\nu},\O) \equiv  \frac{1}{3}
\, \left(\frac{1}{K_1} + \frac{1}{K_2}\right)\,
 \sum_{\a}\,\k(K_{1\a}+K_{2\a})\,\left[{\cal I}_{12}^{\a}+ {\cal J}_{12}^{\a}\right] \,  ,
\ee
and considering that the baryon-to-photon number ratio at recombination is given by $\eta_B \simeq 0.01\,N_{B-L}^{\rm f}$,  one obtains
\be
\eta_B \simeq 0.01 \, \frac{\overline{\ve}(M_1)}{\d_{\rm lep}} \, f(m_{\nu},\O)  ~,
\ee
where the function $f(m_{\nu},\O)$ has quite a complicated dependence on the different parameters ($\theta_{ij}$, $\d$, $\rho$, $\o$). Similar analytical expressions have been given in Ref.~\cite{2RHN} for the hierarchical case. 

\begin{figure}
	\begin{center}
		\includegraphics[width=0.53\textwidth]{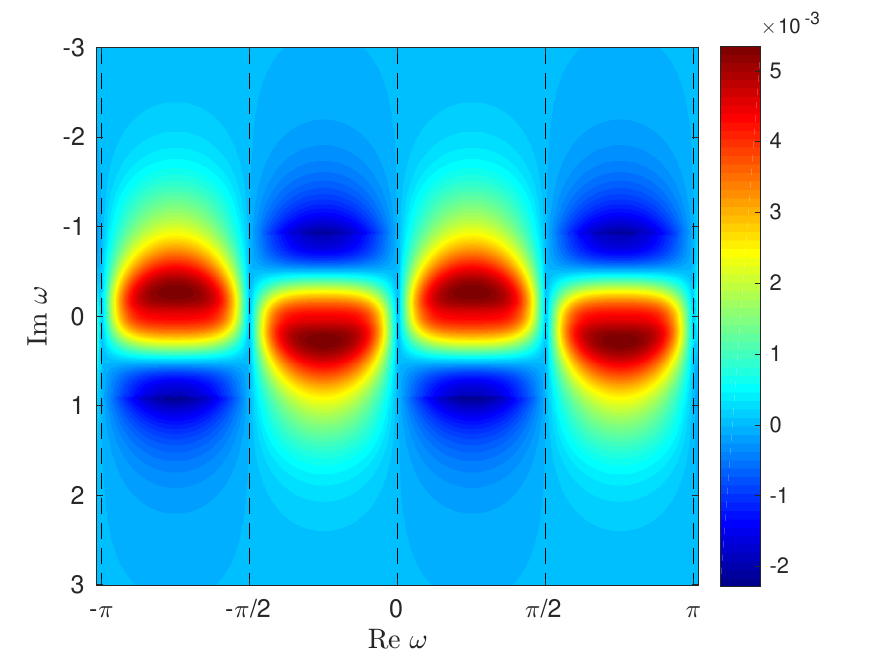}\hspace*{-5mm}
		\includegraphics[width=0.53\textwidth]{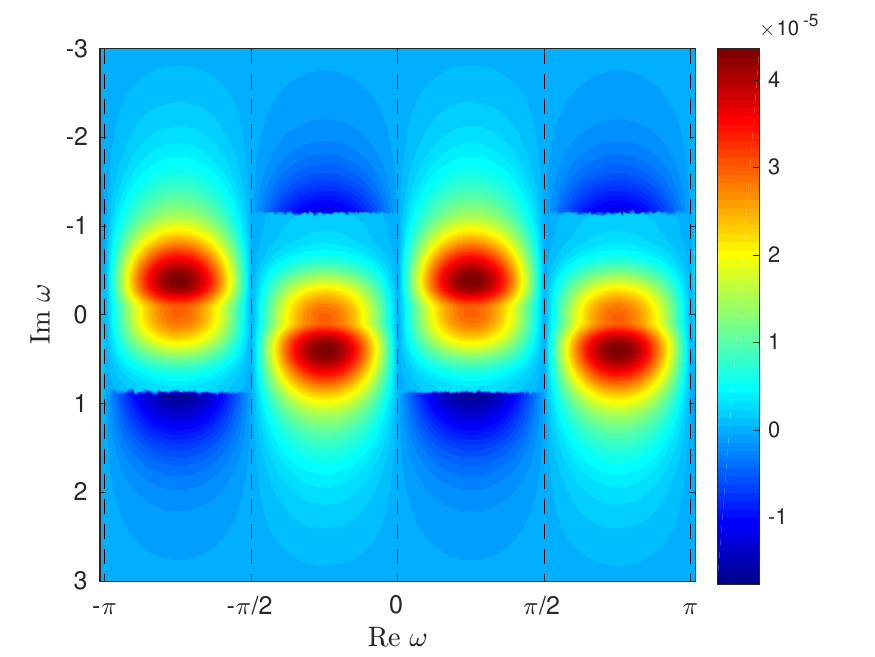}\\
	\end{center}
	\vspace{-5mm}
	\caption{Density plots of the function $f_\mathrm{max}(\omega)$ for NH (left panel) and IH (right panel).}
	\label{fig:f}
\end{figure}

In Fig.~\ref{fig:f}, we show the maximal values
\be
f_\mathrm{max}(\omega) = \mathrm{max}_{\delta,\sigma} \, f(\delta,\sigma,\theta_{ij}^\mathrm{exp},\omega) ~,
\ee
where we have used the best-fit values $\theta_{ij}^\mathrm{exp}$ of Ref.~\cite{global} for the mixing angles. We show density plots for NH (left panel) and IH (right panel), assuming $\z=+1$ (the contour plots for $\zeta=-1$ are obtained from those for $\z = +1$ with the transformation $\o \ra -\o$). The value of $f$ determines the value of $\d_{\rm lep}$ that is needed in order to correctly reproduce the observed asymmetry. Explicitly, for NH (IH),
\be\label{dlep}
\d_{\rm lep} \simeq 0.01 \, \frac{\overline{\ve}(M_1)}{\eta_B}\, f(m_{\nu},\O)
\simeq 0.8 \times 10^{-5} (0.7 \times 10^{-7}) \, \left(\frac{f(m_{\nu},\O)}{f_{\rm max}}\right) \, \left(\frac{M_1}{10^6\, {\rm GeV}}\right)  ~.
\ee
where $f_{\rm max}\simeq 0.005$ ($4 \times 10^{-5}$) is the maximum value of $f$.

\subsubsection*{Interplay between leptogenesis and dark matter results}

This result clearly confirms what we anticipated, that necessarily $M_{\rm DM}=M_3 > M_{\rm S}, M_{\rm I}$. It should also be noticed that since the wash-out is necessarily strong, with $K_{1\a}+K_{2\a} \gtrsim 5$, then $z_{\rm B}(K_{1\a}+K_{2\a}) \gtrsim 4$ and this implies the bulk of the asymmetry to be generated at temperature around $T_{\rm lep} \sim M_{\rm S}/z_{\rm B}$~\cite{pedestrians}. From this, imposing conservatively\footnote{The factor $2$ provides a conservative lower bound, $M_{\rm S}\gtrsim 300\,{\rm GeV}$. When the lower bound is saturated, although $\sim 95\%$ of the lepton asymmetry generated within the interval $[M_{\rm S}/(z_{\rm B}+2),M_{\rm S}/(z_{\rm B}-2)]$ is not converted into a baryon asymmetry, the residual small fraction of the $B-L$ asymmetry generated outside this interval, at higher temperatures $T\sim [140,300]\,{\rm GeV}$, is still converted into a baryon asymmetry and can reproduce the observed asymmetry by further compensating with a value of $\d_{\rm lep}$ lower than the one in Eq.~(\ref{dlep}).} $2\,T_{\rm lep} \gtrsim T_{\rm sph}^{\rm out} \simeq 140$~GeV, one obtains a lower bound, $M_{\rm S} \gtrsim 300$~GeV. This lower bound on $M_{\rm S}$, combined with the Eq.~(\ref{ubMS}) for $M_{\rm DM}^{\rm max}$, for initial vanishing abundance, implies an upper bound $M_{\rm DM}/M_{\rm S} \lesssim 100\,\xi\,\a^{-1/2}_{\rm S}\,\t_{28}^{-1/2}$, which corresponds to $M_{\rm DM}^{\rm max} \lesssim 30\,{\rm TeV}\,\xi\,\a_{\rm S}^{-1/2}\,\t_{28}^{-1/2}$. This upper bound is illustrated by the bottom panel of Fig.~2 (for $\a_{\rm S}=\xi=\tau_{28}=1)$, where we simply plotted $M_{\rm S}$ vs. $M_{\rm DM}$ for different values of $M_{\rm DM}/M_{\rm S}$.

Notice also that in the case of initial vanishing $N_{\rm S}$ abundance, Eq.~(\ref{ubMS}), combined with $M_{\rm DM}/M_{\rm S} \gtrsim 2.2$, implies  the upper bound 
\be
M_{\rm S} \lesssim 8.2 \, \xi^{2/3} \, \a_{\rm S}^{-1/3} \, \t_{28}^{-1/3}~{\rm TeV} \,  ,
\ee
so that the scale of leptogenesis is within the range $T_{\rm lep}={\cal O}(1$--$10\,\xi^{2/3})$~TeV.
In the case of initial thermal $N_{\rm S}$ abundance the Eq.~(\ref{ubthermal}) implies the upper bound
\be
M_{\rm S} \lesssim 330 \, \xi^{4/7} \, \a_{\rm S}^{-5/7} \, \t_{28}^{-2/7} \, {\rm TeV} \,  , 
\ee
so that the scale of leptogenesis is within the range ${\cal O}(1$--$1000\,\xi^{4/7})$~TeV.\footnote{From Eq.~(\ref{hS}) for $h_{\rm S}$, that can be extended to $h_{\rm I}$ by just replacing $\a_{\rm S}$ with an analogous quantity $\a_{\rm I}$, one can see that this corresponds to Yukawas $h_{\rm S}, h_{\rm I} \lesssim 10^{-6},10^{-5} (10^{-5},10^{-4})$ for initial (thermal) $N_{\rm S}$-abundance. This shows that imposing leptogenesis, also in this model one still needs some reduction of the neutrino Yukawa couplings compared to the other massive fermions, although slightly less than in the $\nu$MSM. This reduction should be addressed within some full model, able also to specify the origin of the new interactions.} In this way, we have shown that the model can explain both the DM abundance and the matter-antimatter asymmetry in a unified scenario.\footnote{In this analysis, we have neglected the contribution to the matter-antimatter asymmetry from $N_{\rm S}-N_{\rm I}$ neutrino mixing itself~\cite{ars, teresi, lepmix}. This contribution might relax the constraint on $\d_{\rm lep}$, Eq.~(\ref{dlep}), but has no impact on the obtained constraints from DM and on the properties of the high-energy neutrinos testable at IceCube.}

In Table~\ref{tab:results} we summarise the results for the allowed window on $M_{\rm DM}$ both for the quasi-degenerate and for the hierarchical case (imposing also successful leptogenesis) for the same values of $M_{\rm DM}/M_{\rm S}$ as in Fig.~2.

\begin{table}
	\centering
\begin{tabular}{|c|c|c|c|c|c| }
				      \hline
\backslashbox{$\xi$}{$M_{\rm DM}/M_{\rm S}$} &  $\simeq 1$   &   2.3 &  10 & $100$ & 1000   	\\
                    \hline
                        1 & NON VIABLE &  $M^{\star}_{\rm DM}\simeq 8\,{\rm TeV}$ & $(3$---$14)\,{\rm TeV}$  & $M^{\star}_{\rm DM}\simeq 30\,{\rm TeV}$ & NON VIABLE\\
                        \hline
                        10 &  $M^{\star}_{\rm DM} \simeq 30\,{\rm TeV}$ & $(0.7$---$40)\,{\rm TeV}$ & $($3---$63)\,{\rm TeV}$ 
                        & $(30$---$135)\,{\rm TeV}$ &  $M_{\rm DM}^{\star} \simeq 300\, {\rm TeV}$ \\
                    \hline
                    \end{tabular} 
			\caption{Summary of the results for the allowed $M_{\rm DM}$ window in the case of initial vanishing abundance and when successful leptogenesis is imposed (which requires $M_{\rm S} \gtrsim 300\,{\rm GeV}$) for $\tau^{\rm min}_{\rm DM}=10^{28}\,{\rm s}$, $\a_{\rm S}=1$ and for the indicated values of $M_{\rm DM}/M_{\rm S}$ and $\xi$. We recall that the critical value $M^{\star}_{\rm DM}$ corresponds to the case when the window reduces to a point ($M^{\star}_{\rm DM} \equiv M_{\rm DM}^{\rm max}=M_{\rm DM}^{\rm min}$). For $\xi =1$, the minimal allowed value of $M_{\rm DM}/M_{\rm S}$ is given by $M_{\rm DM}/M_{\rm S}\simeq 2.3$ and the maximal value by $M_{\rm DM}/M_{\rm S}\simeq 100$. For the quasi-degenerate case, an allowed window starts to open up for $\xi \simeq 10$ at $M_{\rm DM}^{\star}\simeq 30\,{\rm TeV}$, which corresponds to $\d_{\rm DM}^{\rm min}\simeq 0.1$.}
		\label{tab:results}
\end{table}

In Table~\ref{tab:results2} we summarise  the allowed windows for the different relevant quantities of the model as indicated for the case of initial vanishing abundance and for $\xi =1, 10$.

\begin{table}
	\centering
\begin{tabular}{|c|c|c|}
				      \hline
		&  $\xi = 1$   &   $\xi = 10$    \\
        \hline
  $M_{\rm DM}/M_{\rm S}$           &   $2.2 \,$ ---$\, 100 $                 &      $1.07\,$---$10^3$             \\
        \hline
  $M_{\rm DM}$                             &    $1.3\,{\rm TeV}$---$30\,{\rm TeV}$    &    $(0.45\,$---$\,300)\,{\rm TeV}$               \\
        \hline    
  $z_{\rm res} $                            &  $(1.5\,$---$\,5) \times 10^{-7}$                 &       $1.5 \times 10^{-7}\,$---$\, 4 \times 10^{-6}$             \\
        \hline          
   $T_{\rm res}$                             &   $(1\times 10^{10}\, $---$\, 2 \times 10^{12})\,{\rm GeV}$    &       $(3\times 10^{9}\,$---$\,2\times 10^{12})\,{\rm GeV}$            \\
        \hline          
    $T_{\rm lep}\equiv M_{\rm S}/z_B$                       &        $140\,{\rm GeV}\,$---$\,1.75\,{\rm TeV}$            &   $140\,{\rm GeV}\,$---$\,15\,{\rm TeV}$              \\
        \hline                       
\end{tabular} 
			\caption{Allowed windows, imposing successful DM+leptogenesis, for the indicated parameters in the case of initial vanishing abundance  for $\xi=1,10$.}
		\label{tab:results2}
\end{table}

From these two tables one can see how, in the case of a mild hierarchy $M_{\rm DM}/M_{\rm S} \lesssim 10$, one has (for $\xi=1$) $M_{\rm S} \lesssim 1.4\,{\rm TeV}$ and $M_{\rm DM}\lesssim 14\,{\rm TeV}$. In this way, one can think of containing all new physics below ${\cal O}(10\,{\rm TeV})$, so that electroweak scale stability can be obtained with a reasonable fine-tuning without the necessity of resorting to specific solutions, such as a supersymmetric extension, to the naturalness problem. On the other hand, for $M_{\rm DM}/M_{\rm S} \gtrsim 10$ one introduces a new very high energy scale, above ${\cal O}(10\,{\rm TeV})$, associated to $M_{\rm DM}$ and, if one wants to address naturalness, an extension, such as supersymmetry, would be desirable. In this case the discussed constraints would get modified along similar lines  extensively studied already  in the case of leptogenesis~\cite{susylep}. The modifications could be again parameterised in terms of a contribution to the parameter $\xi$, that in the case of supersymmetry would likely be quite mild ($\xi_{\rm SUSY} \sim {\cal O}(1)$).

\section{High-energy neutrinos from DM decays and IceCube data}
\label{sec:IceCube}

In this section, we finally discuss the contribution to the very high-energy neutrino flux from DM decays and its properties, comparing the predictions with the most recent IceCube high-energy starting event (HESE) data~\cite{Aartsen:2015zva}. As we discussed in Section~\ref{sec:DMdecays}, $N_{\rm DM}$ dominantly decays through two-body and four-body processes (see Fig.~\ref{fig:decaychannels}).  In this section we consider two-body decays, which occur via the mixing of $N_{\rm S}$ with leptons, Higgs and gauge bosons, with ratio of branching ratios at the source~\cite{pascoli}, $(f_e : f_\mu : f_\tau)_{\rm S}$,
\be
\left( BR(N_{\rm S} \ra {\ell}^{\pm}\,W^{\mp}) : BR(N_{\rm S}\ra \nu_{\a} Z,\bar{\nu}_\a Z) : BR(N_{\rm S} \ra H\n_\a,H\bar{\n}_\a) \right)_{\rm S} = (2:1:1)_{\rm S}  ~.
\ee

We first discuss the flavour composition of the (almost monochromatic) neutrinos directly produced by $N_{\rm S}$ decays, with its distinctive features, and next we derive the event energy spectrum showing two representative choices of $(\tau_{\rm DM},M_{\rm DM})$ plus an astrophysical power-law flux, that altogether predict an spectrum in good agreement with the 4-year IceCube data.

\subsection{Flavour composition of hard neutrinos}

Whereas gauge bosons and Higgs decays generate a softer neutrino flux with $(1:1:1)_{\rm S}$ flavour composition at production, neutrinos produced directly from $N_{\rm S}$ decays retain information of the Yukawa structure of the model (see Eq.~(\ref{WASS})) and thus, are of particular interest. These hard neutrinos are also generated from four-body decays, $N_{\rm DM} \ra N_{\rm S} + 2\,A \ra 3\,A+ \nu_{\rm S}$, but in this case, their relevance is further diluted compared to the softer neutrinos. They have the highest kinematically allowed energies $M_{\rm DM}/2$, so they are expected to produce the events from DM decays with the highest energies. Therefore, statistics permitting, by analysing the events close to the high-energy endpoint of the spectrum, information about the flavour composition and thus, about the Yukawa structure, might be inferred. However, let us note that depending on the type of interaction and on the flavour of the incoming neutrino, the deposited energy in the detector (which is the current observable used by the IceCube collaboration) might be quite different from the actual neutrino energy. Indeed, only for electron neutrino and antineutrino charged-current interactions off nucleons, the energy deposited in the detector is close to the neutrino energy. This makes the discrimination of the contribution from the neutrinos at the kinematical threshold a very challenging task, certainly not possible with current data.  However, despite these intrinsic experimental difficulties in detecting these (almost monochromatic) neutrinos, we still think it is interesting to discuss their flavour composition, showing that, at production, it can be quite different from the standard mechanisms. Let us then first discuss their flavour composition at production and then at Earth.

\subsubsection{Flavour composition of hard neutrinos at production}

The flavour composition of monochromatic neutrinos at production is determined by the $N_{\rm S}$-flavour branching ratios ($\a=e,\m,\t$)
\begin{equation}\label{ratio_source}
f_{\a, \rm S} \equiv \frac{\G_{i \alpha}}{\G_{\rm S}} = \frac{|m_{D \alpha i}|^2}{(m^{\dagger}_D\,m_D)_{ii}} ~,
\end{equation}
where $i=1$ or $2$ is the index corresponding to $N_{\rm S}$.  Let us now  express the $f_{\a}$'s in terms of the low-energy neutrino parameters using the convenient orthogonal parameterisation. Taking for $m_D$ the third form in Eq.~(\ref{WASS}), recalling that $N_{\rm DM}=N_3$, from the orthogonal parameterisation in Eq.~(\ref{orthogonal1}), one straightforwardly finds
\begin{equation}
f_{\a,\rm S} = \frac{|U_{\a 2} \, \sqrt{m_2} \, \O_{2i} + U_{\a 3} \, \sqrt{m_3} \,\O_{3i}|^2}{m_2 \, |\O_{2i}|^2 + m_3 \, |\O_{3i}|^2} ~.
\end{equation}

\begin{figure}
	\begin{center}
		\includegraphics[width=0.45\textwidth]{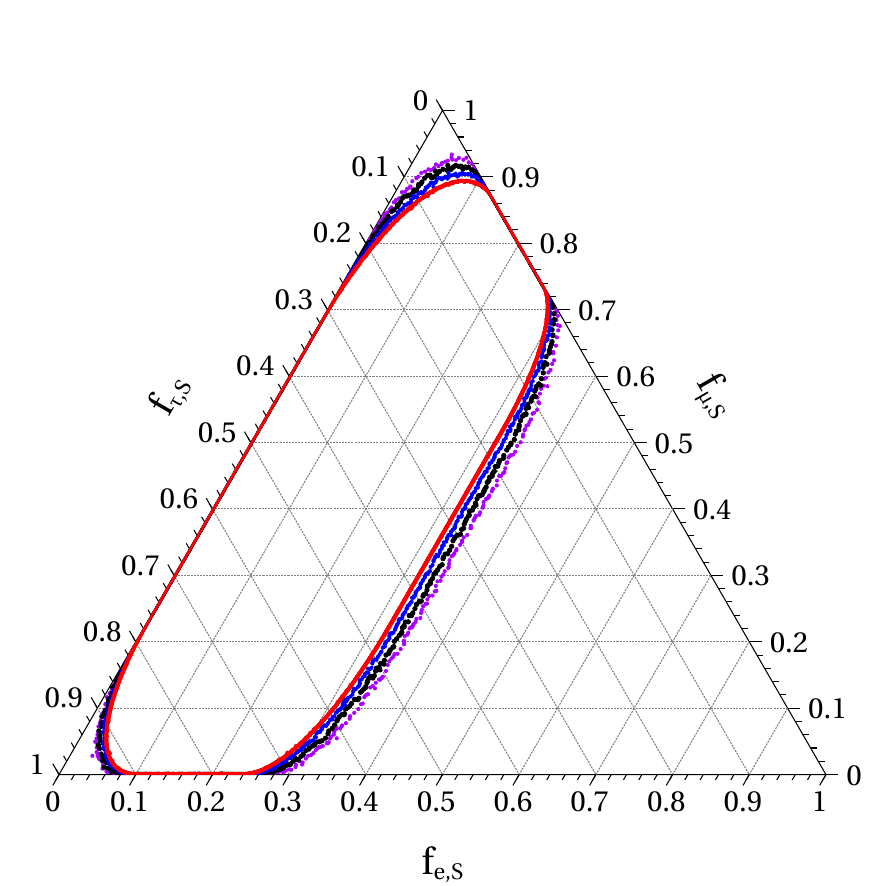}\hspace*{5mm}
		\includegraphics[width=0.45\textwidth]{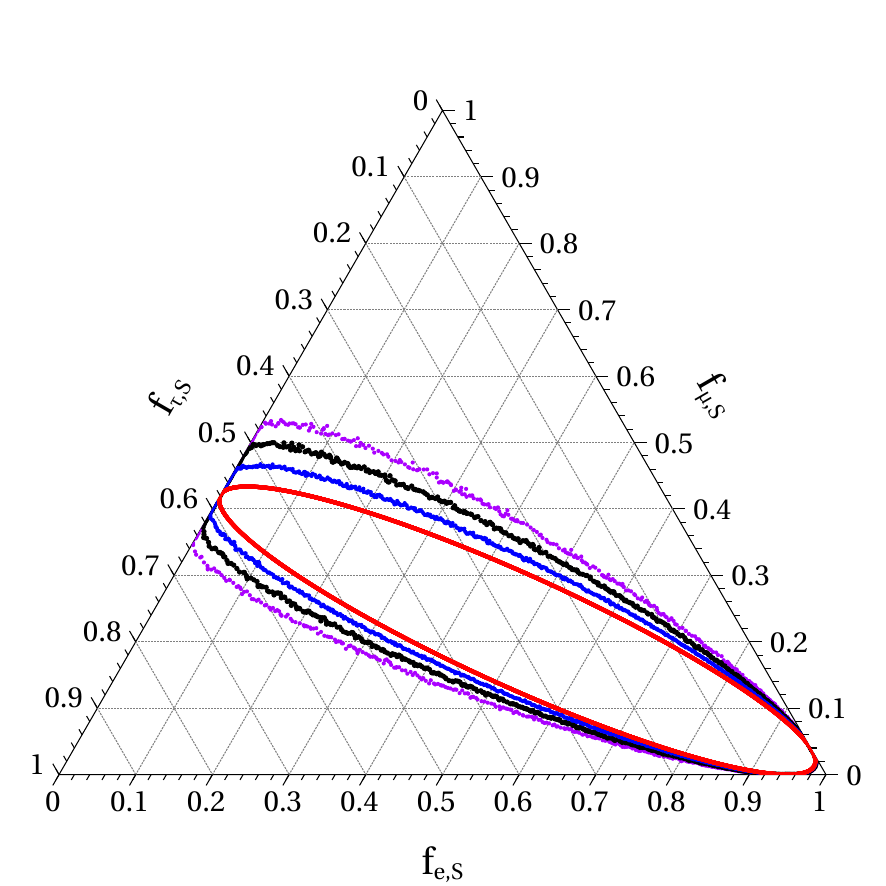}\\
	\end{center}
	\vspace{-5mm}
	\caption{Flavour composition at the source of the hard neutrinos from DM decays. We show the results for the best-fit values of the mixing parameters (red curves), $1\s$ C.L. (blue curves), $2\s$ C.L. (black curves) and $3\s$ C.L. (purple curves), for NH (left panel) and for IH (right panel).}
	\label{fig:source}
\end{figure}

Notice that the denominator is $\widetilde{m}_{\rm S}$. This expression clearly implies the upper bound
\begin{equation}
f_{\alpha,\rm S} \leq  |U_{\alpha 2}|^2 + |U_{\alpha 3}|^2 = 1 - |U_{\alpha 1}|^2  \,   .
\end{equation}
Using the $3\sigma$ C.L. lower bounds on the mixing matrix elements of the global fit of~\cite{global}, one obtains
\begin{equation}
\label{upperbounds}
\begin{split}
& f_{e,\rm S} \lesssim 0.36,\quad f_{\mu,\rm S} \lesssim 0.95, \quad f_{\tau,\rm S} \lesssim 0.94 \quad 
\;\;\; \text{(NH)},\\
& f_{e,\rm S} \lesssim 0.98,\quad f_{\mu,\rm S} \lesssim 0.62, \quad f_{\tau,\rm S} \lesssim 0.65 \quad 
\;\;\; \text{(IH)} ~.
\end{split}
\end{equation}
In the left (right) panel of Fig.~\ref{fig:source}, we show the allowed regions at $1\s$ C.L., $2\s$ C.L. and $3\s$ C.L. (according to the $\chi^2$-projections provided in Refs.~\cite{global,nufit}) for the flavour fractions at the source, $f_{\a,\rm S}$, for NH (IH).  One can see that, as expected, these plots respect the analytical bounds found in Eqs.~(\ref{upperbounds}).

In Section~\ref{sec:event_spectrum}, as our benchmark scenario we consider the flavour composition of these hard neutrinos to be $(0:1:1)_{\rm S}$.

\subsubsection{Flavour composition of hard neutrinos at Earth}

\begin{figure}
        \begin{center}
                \includegraphics[width=0.45\textwidth]{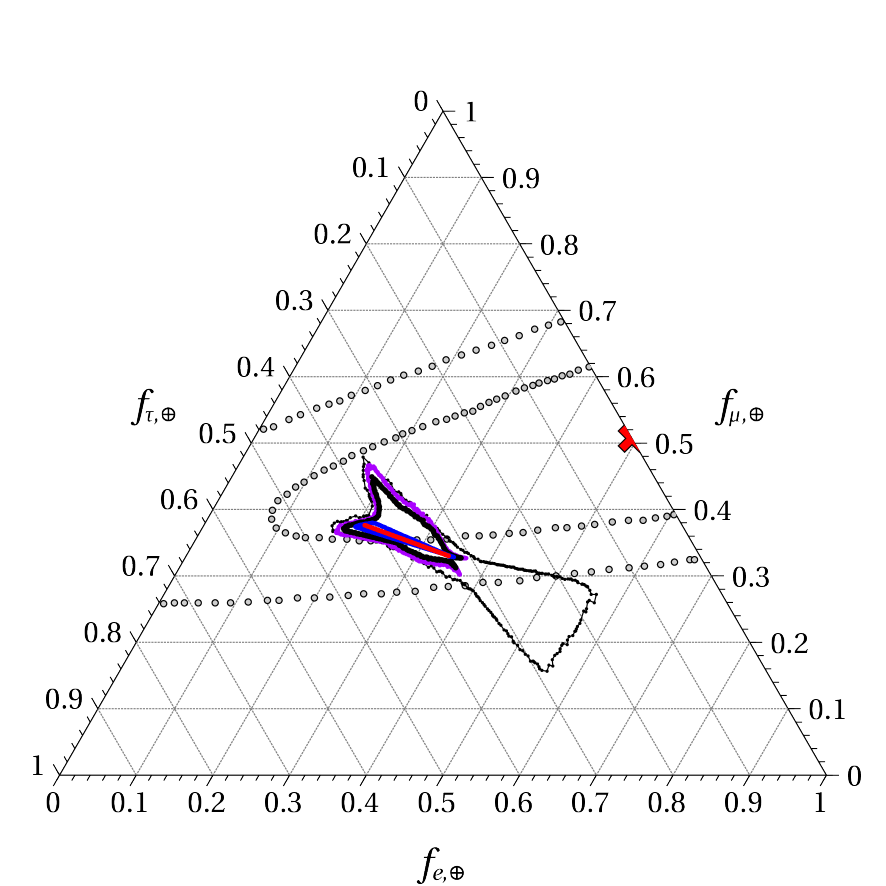}\hspace*{5mm}
                \includegraphics[width=0.45\textwidth]{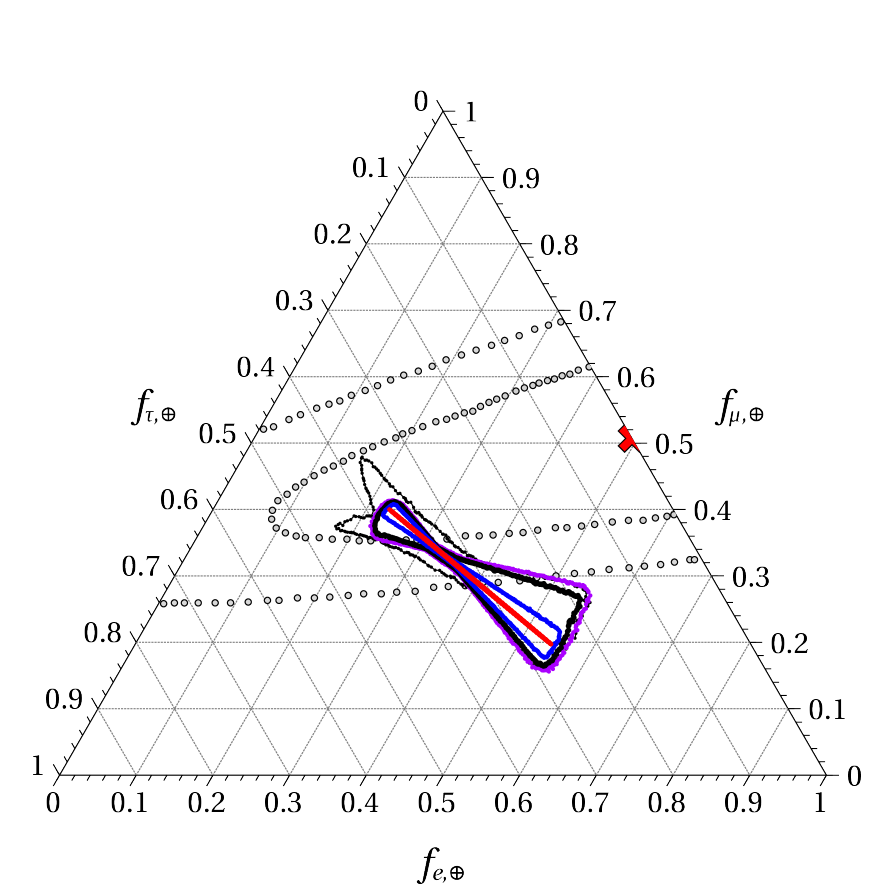}
        \end{center}
        \vspace{-5mm}
        \caption{Flavour composition at Earth for NH (left panel) and IH (right panel) with the same color coding as in Fig.~\ref{fig:source}. The red cross is the IceCube best fit point and the areas bounded by the grey lines are the 68\% and 95\% confidence regions, respectively~\cite{ICflavour}.}
        \label{fig:detector}
\end{figure}

In order to compute the final flavour composition at the detector, for all the fluxes in this work, we assume standard neutrino oscillations without exotic interactions. In this case, the neutrino flavour states produced at the source are subject to averaged oscillations in their way to the Earth. Therefore, the probability for a neutrino to arrive at the detector with flavour $\alpha$, if it was produced with flavour $\beta$, is given by
\begin{equation}
P_{\alpha\beta} = \sum_j |U_{\alpha j}|^2 |U_{\beta j}|^2 ~,
\end{equation}
and correspondingly the flavour composition at Earth, in terms of the flavour composition at the source, is given by
\begin{equation}
f_{\alpha,\oplus} = \sum_\beta P_{\alpha\beta} f_{\beta,\rm S} \,  .
\end{equation}
The flavour composition at Earth of the monochromatic neutrinos from DM decays is shown in Fig.~\ref{fig:detector}. Looking at the plot for NH (left panel), we find the main constraint on the flavour composition to be in the electron component, which is restricted to be $f_{e,\oplus} \lesssim 1/3$. This feature can be qualitatively understood in the following way. Using the fact that $f_{e,\rm S}+f_{\mu,\rm S}+f_{\tau,\rm S}=1$ and the normalization of $P_{\alpha\beta}$, i.e., \ $\sum_\beta P_{\alpha\beta} = \sum_\alpha P_{\alpha\beta}=1$, one can derive the following expression for the electron flavour content at the detector:
\begin{equation}
f_{e,\oplus} = P_{e\tau} + f_{e,\rm S} \left( P_{ee} - P_{e\tau }\right) + f_{\mu,\rm S} \left( P_{e\mu}-P_{e\tau} \right) ~,
\end{equation}
where $P_{e\tau}\approx 1/5$, $P_{ee} - P_{e\tau}\approx 1/3$ and $|P_{e\mu}-P_{e\tau}| \approx 0.05$. Thus, the total electron component at the detector is quite insensitive to the muon component at the source, and we arrive at the approximate relation
\begin{equation}
f_{e,\oplus} \approx \frac{1}{5} + \frac{f_{e,\rm S}}{3} ~,
\end{equation}
i.e., the electron component at the detector can be between about 0.2 and 0.6. Our model predicts for NH $f_{e,\rm S}\lesssim 1/3$, i.e., also $f_{e,\oplus}\lesssim 1/3$. For IH the restrictions are much less pronounced, as can be seen from Fig.~\ref{fig:detector} (right panel). 

\begin{figure}[t]
	\begin{center}
		\includegraphics[width=0.45\textwidth]{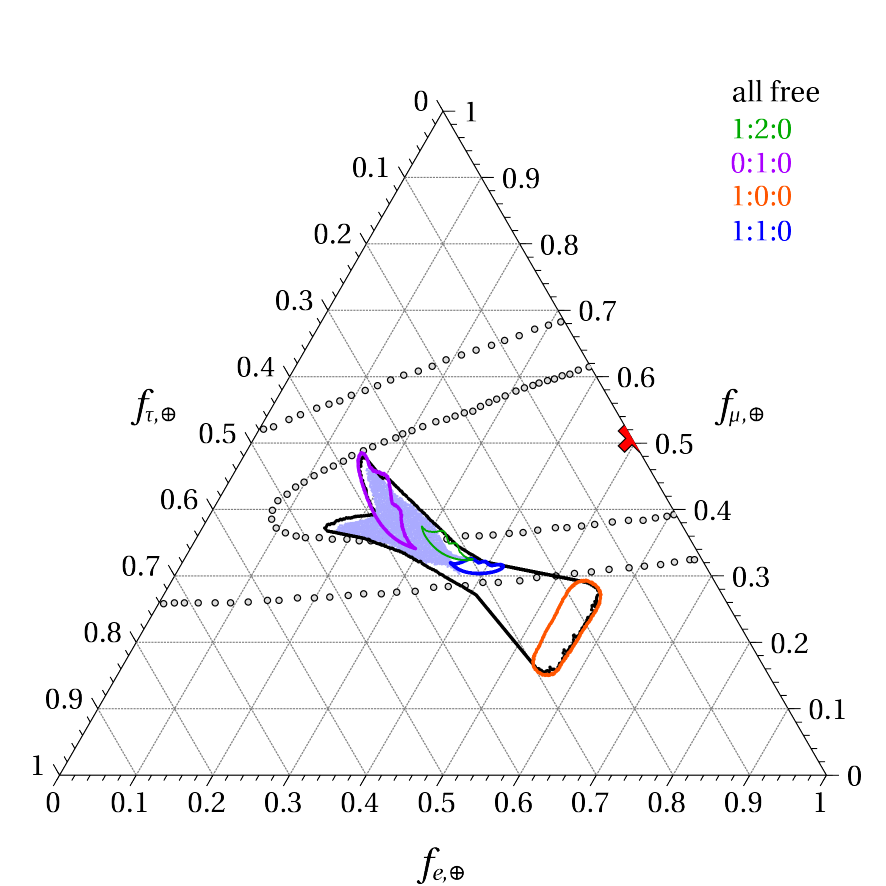}\hspace*{5mm}
		\includegraphics[width=0.45\textwidth]{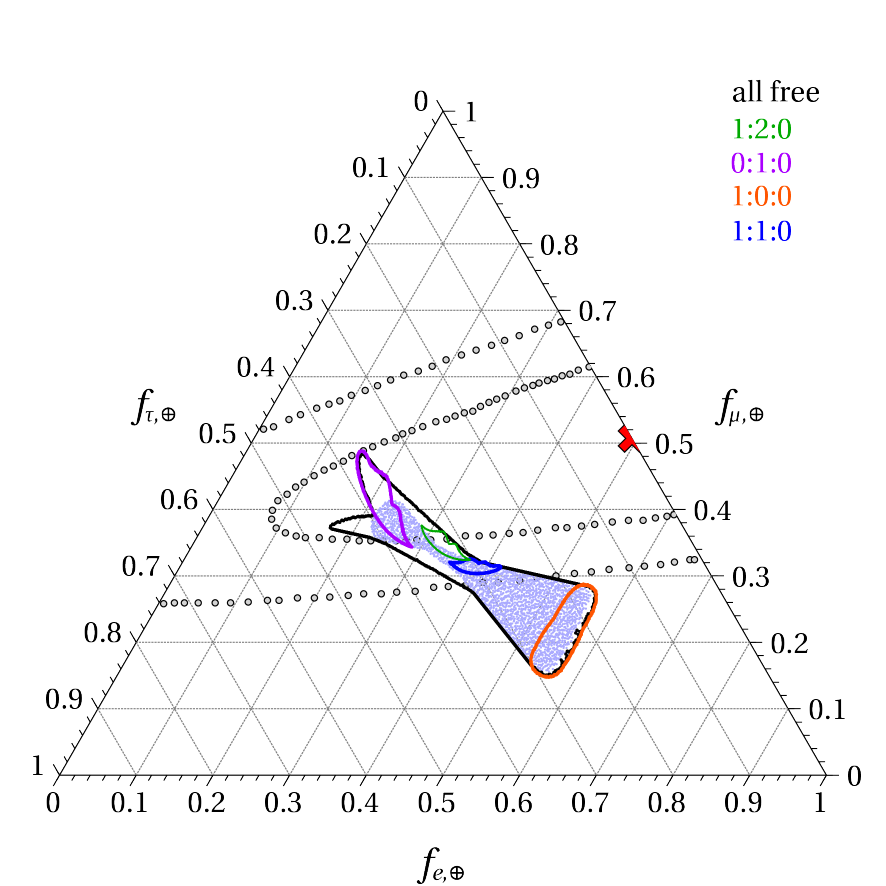}
	\end{center}
	\vspace{-5mm}
	\caption{Flavour composition at Earth of the hard neutrinos at the high-energy end of the DM decay spectrum in our model (light blue filled area) and four reference cases for normal (left panel) and inverted (right panel) hierarchical spectrum. The large black area is the maximal allowed area for arbitrary flavour composition at the source. In order to produce this plot, the $3\s$ C.L. ranges of the neutrino oscillation parameters have been used. The red cross is the IceCube best fit point and the areas bounded by the grey lines are the 68\% and 95\% confidence regions, respectively~\cite{ICflavour}.}
	\label{fig:comparison_standard}
\end{figure}

Finally, in Fig.~\ref{fig:comparison_standard}, we compare the predictions of our model for the flavour composition at detection of the monochromatic neutrinos at the high-energy end of the DM decay spectrum to some of the standard reference cases. The flavour compositions at the source of the cases we consider are: $(f_{e}:f_{\mu}:f_{\tau})_{\rm S} = (1:2:0)_{\rm S}$ (pion beam), $(0:1:0)_{\rm S}$ (muon damped), $(1:1:0)_{\rm S}$ and $(1:0:0)_{\rm S}$ (neutron decay).

Finally, let us stress again that the ranges of the flavour composition shown in Figs.~\ref{fig:source}-\ref{fig:comparison_standard} only correspond to the almost monochromatic flux at the high-energy end of the DM decay spectrum. Given the relatively hard neutrino spectrum produced from gauge boson decays and the IceCube detector capabilities, it will be extremely challenging to single out this contribution. On the other hand, this could offer important information about the specific Yukawa structure, a definite feature of the model.

\subsection{Event energy spectrum}\label{sec:event_spectrum}

Here we consider two representative cases for the DM mass and lifetime and compute the event spectra from two-body DM decays, as described above, expected after 4 years of data taking in IceCube. Since the DM signal alone does not represent a good fit to the entire data sample, we also consider an astrophysical contribution with a power-law flux.

The flux from DM decays has two contributions: galactic and extragalactic, 
\begin{equation}
\frac{d\Phi_{\rm DM}}{dE_\nu}  = \frac{d\Phi_{\rm DM}^{\rm xg}}{dE_\nu} + \frac{d\Phi_{\rm DM}^{\rm g}}{dE_\nu} ~.
\end{equation}
The decays of DM particles at cosmological distances produce a nearly isotropic neutrino and antineutrino flux, which is given by
\begin{equation}
\frac{d\Phi_{\rm DM}^{\rm xg}}{dE_\nu} = \frac{\Omega_{\rm DM} \, \rho_{\rm c, 0}}{4 \, \pi \, M_{\rm DM} \, \tau_{\rm DM}} \, \int_0^\infty \frac{1}{(1+z) \, H(z)} \, \frac{dN_\nu}{dE_\nu} [(1+z) \, E_\nu] \, dz ~,
\label{eq:xgal}
\end{equation}
where $\Omega_{\rm DM} = 0.2618$ is the fraction of DM density today, $H(z) = H_0 \, \sqrt{\Omega_\Lambda + \Omega_{\rm m} \, (1+z)^3}$ is the Hubble expansion rate as a function of redshift, with $\Omega_\Lambda = 0.6879$ and $\Omega_{\rm m} = 0.3121$~\cite{planck} (best fit values). The neutrino energy spectrum (of each flavour) from DM decays, $dN_\nu/dE_\nu$, depends on the DM decay channel and on the DM mass, but in our notation we omit to make these dependences explicit. We use the tabulated results in Ref.~\cite{Cirelli:2010xx}, which include electroweak corrections~\cite{Ciafaloni:2010ti} and were computed using PYTHIA 8.135~\cite{Sjostrand:2007gs}, and are provided for annihilations of DM particles with mass up to $M_{\rm DM} = 100$~TeV (or DM decays up to $M_{\rm DM} = 200$~TeV) and rescale them for higher masses, as done in Ref.~\cite{serpico}. Let us note that this procedure tends to slightly overestimate the final flux, although the precise factor depends on the decay channel and extrapolated value of the DM mass. Thus, our quoted values for the DM lifetime would need to be correspondingly scaled down by the same factor.

In addition to the extragalactic signal, DM decays in the Milky Way would also produce a flux of neutrinos and antineutrinos, which would be higher in the Southern hemisphere. Unlike the extragalactic flux, the shape of the galactic flux is not distorted by the redshifting of the signal, which implies differences also in the energy spectrum. The neutrino and antineutrino flux in a direction with galactic coordinates $(l,b)$ produced by DM decays in our own halo is given by
\begin{equation}
\frac{d\Phi_{\rm DM}^{\rm g}}{dE_\nu} [l,b] = \frac{1}{4 \, \pi \, M_{\rm DM} \, \tau_{\rm DM}} \, \frac{dN_\nu}{dE_\nu} \, \int_0^\infty \rho[r(s,l,b)] \, ds  ~,
\label{eq:gal}
\end{equation}	
where $\rho(r)$ is the DM density profile of the Milky Way as a function of the distance from the galactic centre, $r$. For a given distance over the line-of-sight, $s$, the galactocentric distance depends on the galactic longitude, $l$ and the galactic latitude, $b$, as
\begin{equation}
r(s,l,b) = \sqrt{s^2 + R_\odot^2 - 2 s R_\odot \cos{b} \cos{l}} ~, 
\label{eq:GCdistance}
\end{equation}
where $R_\odot = 8.33$~kpc is the distance from the Sun to the galactic centre~\cite{Gillessen:2008qv}.

For the DM distribution in our galaxy we adopt a generalised Navarro-Frenk-White~\cite{genNFW} density profile,
\begin{equation}
\rho(r) = \frac{\rho_s}{(r/r_s)^\gamma \, (1 + r/r_s)^{3-\gamma}} ~, 
\end{equation}
with a scale radius $r_s = 20$~kpc, $\gamma = 0.75$ and a local DM density $\rho(R_\odot) = 0.42$~GeV/cm$^3$~\cite{Pato:2015dua}. Let us note, however, that the linear dependence on the density of the flux from DM decays, unlike the quadratic dependence of the flux from DM annihilations, implies smaller uncertainties from the poorly known shape of the DM profile and thus, the precise choice of the DM distribution in the halo is less relevant. In any case, these uncertainties would affect the normalization and the angular dependence of the flux, but not the overall shape of the energy spectrum of the galactic signal.

\begin{figure}[t]
	\begin{center}
		\includegraphics[width=0.55\textwidth]{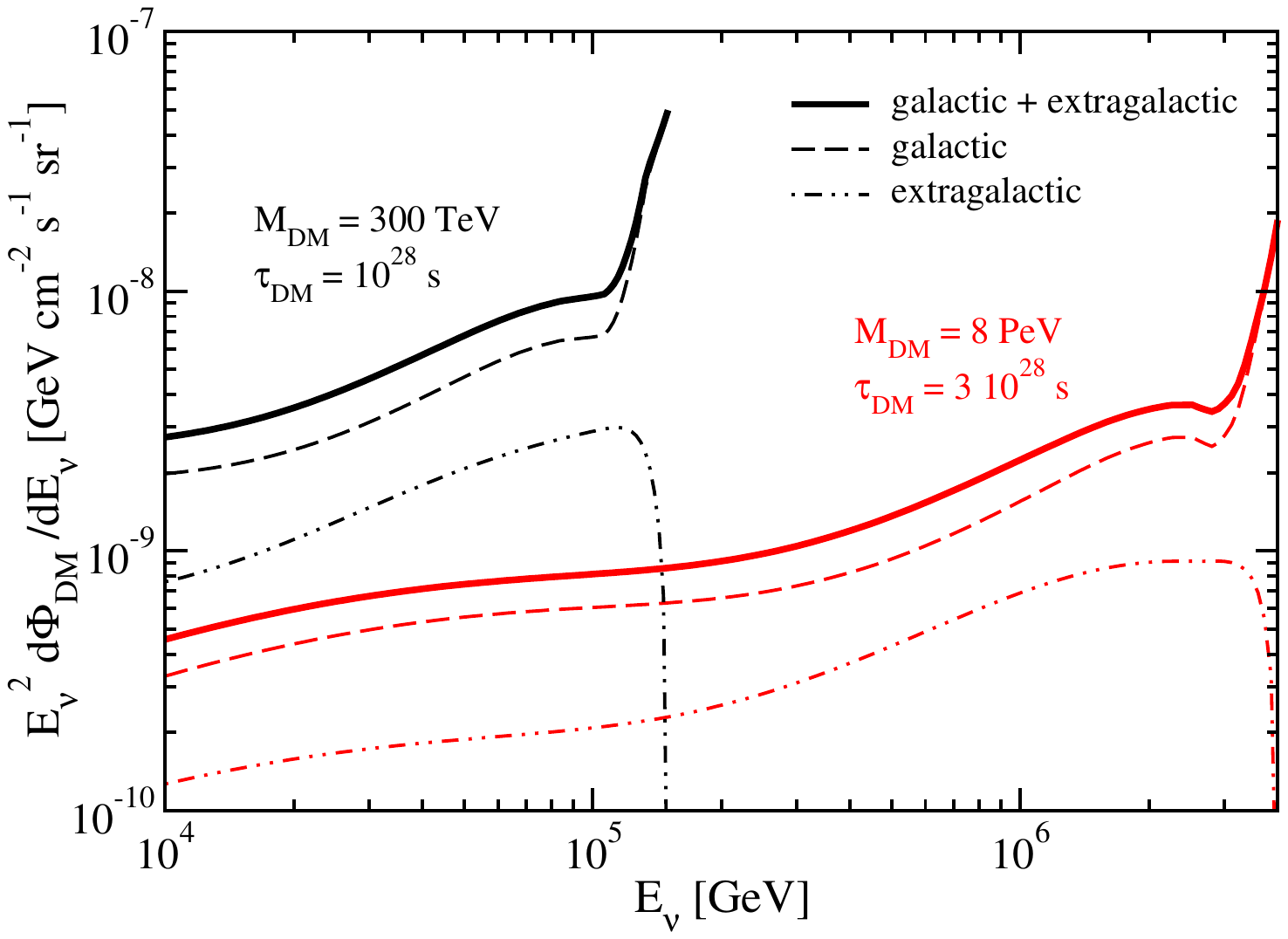}\hspace*{-7mm}
		\includegraphics[width=0.55\textwidth]{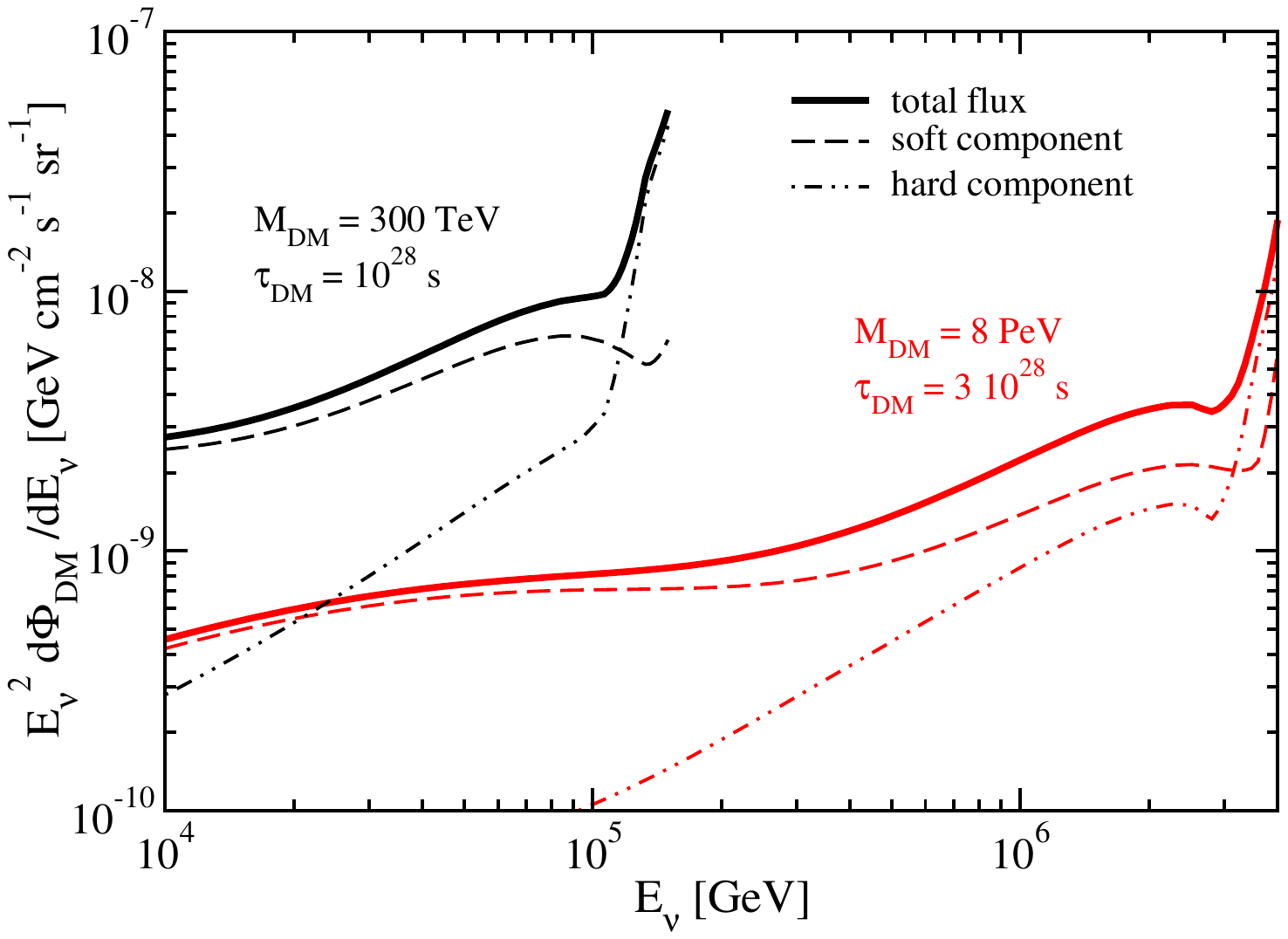}
	\end{center}
	\vspace{-5mm}
	\caption{Flavour-averaged neutrino flux, $(\nu_e+\nu_\mu+\nu_\tau)/3$, at Earth from DM decays for $M_{\rm DM} = 300$~TeV and $\tau_{\rm DM} = 10^{28}$~s (black curves) and $M_{\rm DM} = 8$~PeV and $\tau_{\rm DM} = 3 \times 10^{28}$~s (red curves). Left panel: Galactic (dashed curves) and extragalactic (dot-dashed curves) contributions. Right panel: soft component (dashed curves) and hard component (dot-dashed curves). In both panels, we show the total flux (solid curves).}
	\label{fig:fluxes}
\end{figure}

In Fig.~\ref{fig:fluxes}, assuming NH and the best fit values in Eqs.~(\ref{eq:expranges}), we show the flavour-averaged neutrino flux, $(\nu_e+\nu_\mu+\nu_\tau)/3$, at Earth after propagation for two different DM masses and lifetimes: $M_{\rm DM} = 300$~TeV and $\tau_{\rm DM} = 10^{28}$~s (black curves) and $M_{\rm DM} = 8$~PeV and $\tau_{\rm DM} = 3 \times 10^{28}$~s (red curves). 
\footnote{From Table 1 one can see that, for initial vanishing $N_{\rm S}$ abundance, the first case can explain both dark matter and leptogenesis if $\xi=10$, one has then to resort to some of the cases discussed in Section 2.4. 
The second case would on the other hand require values $\xi={\cal O}(1000)$, too large to be realistic. Both cases can explain just dark matter either, in the first case,  with $\xi=1$ or, in the second case, with a plausible value $\xi \gtrsim 10$. From
this point of view the model certainly prefers the first case, i.e., solutions with $M_{\rm DM}\sim {\cal O}$(100\,{\rm TeV}).
Of course for initial thermal $N_{\rm S}$ abundance both cases would be able to explain both dark matter and matter-antimatter
asymmetry of the universe with $\xi=1$.}
In the left panel we depict the galactic (dashed curves) and extragalactic (dot-dashed curves) contributions, as well as the total flux (solid curves), whereas in the right panel we show the soft component (dashed curves), i.e., neutrinos from gauge bosons, Higgs and leptons decays and from the related electroweak corrections, and the hard component (dot-dashed curves), i.e., neutrinos produced at the decay vertex (including the related electroweak corrections). We note that the galactic flux dominates over the extragalactic contribution and that the soft component of the flux dominates over the hard one, except at the highest energies, close to the kinematical threshold.

Finally, we also compute the event energy spectra for these fluxes and compare them with the 4-year IceCube HESE data. As we note below, the DM decay signal cannot explain all the observed events, so another component of the flux is required. Therefore, in addition to the events from DM decays, we add the contribution from an astrophysical flux described by a power-law spectrum, 
\begin{equation}
\frac{d\Phi_{\rm a}}{dE_\nu} = \phi \, \left(\frac{E_\nu}{100 \, {\rm TeV}}\right)^{-\gamma} ~,
\label{eq:astro}
\end{equation}
where $\phi$ is the normalization of the flux, in units of $10^{-18} \, {\rm GeV}^{-1} \, {\rm cm}^{-2} \, {\rm s}^{-1} \, {\rm sr}^{-1}$. For this astrophysical neutrino flux we assume the canonical flavour composition at Earth, $(1:1:1)_\oplus$.

\begin{figure}[t]
	\begin{center}
		\includegraphics[width=0.76\textwidth]{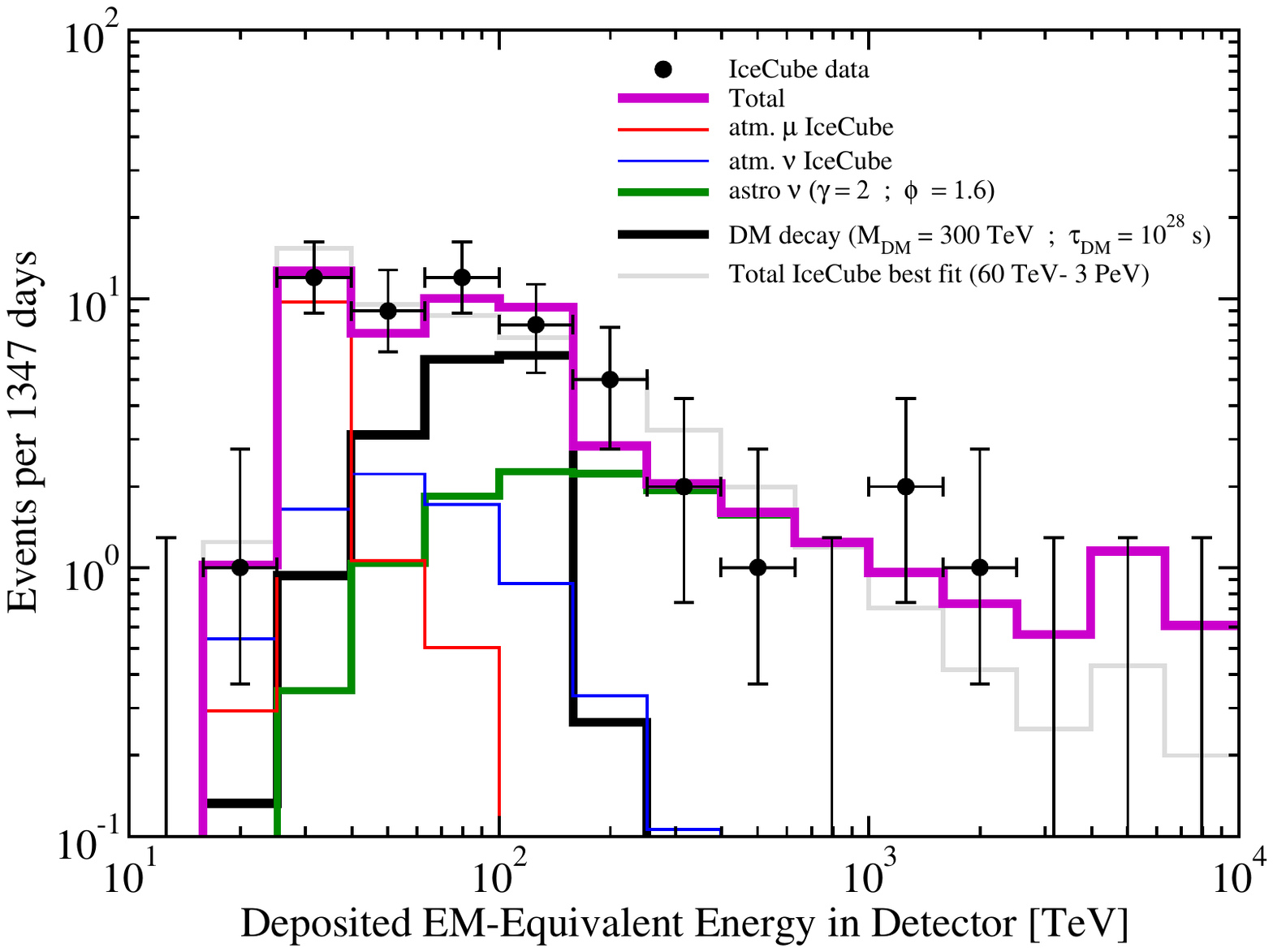} \vspace{-5mm}
	\end{center}
	\caption{Event spectra in the IceCube detector after 1347 days: atmospheric muon events (red histogram); conventional atmospheric neutrino events (blue histogram); astrophysical neutrino events (green histogram), $E_\nu^2 \, d\Phi_{\rm a}/dE_\nu = 1.6 \times 10^{-8} \, {\rm GeV} \, {\rm cm}^{-2} \, {\rm s}^{-1} \, {\rm sr}^{-1}$; events from DM decays (black histogram), $M_{\rm DM} = 300$~TeV and $\tau_{\rm DM} = 10^{28}$~s; and total event spectrum (purple histogram). We also show the spectrum obtained using the preliminary IceCube best fit for $(1:1:1)_\oplus$ in the EM-equivalent deposited energy interval [60~TeV $-$ 3~PeV] (gray histogram), $E_\nu^2 \, d\Phi/dE_\nu = 6.6 \times 10^{-8} \, (E_\nu/100 \, {\rm TeV})^{-0.58}  \, {\rm GeV} \, {\rm cm}^{-2} \, {\rm s}^{-1} \, {\rm sr}^{-1}$, and the binned high-energy neutrino event data (black dots)~\cite{Aartsen:2015zva} with Feldman-Cousins errors~\cite{Feldman:1997qc}.}
	\label{fig:events300}
\end{figure}

In this work we do not attempt to perform a fit to this combined model (DM decays plus power-law flux), but simply to show some exemplary cases (see, however, Ref.~\cite{BEPS}). Therefore, after choosing some representative values for the DM parameters, we fix the normalization of the astrophysical flux by imposing the total number of events in the electromagnetic(EM)-equivalent deposited energy range $[60~{\rm TeV}$---$10~{\rm PeV}]$ to be equal to the sum of the DM decay and astrophysical signals plus the expected backgrounds. For the atmospheric muon and neutrino backgrounds we scale the 3-year (988 days) IceCube expected numbers~\cite{icecube} to obtain the 4-year (1347 days) expectations, i.e., we consider 3.3 atmospheric neutrino events and 0.6 atmospheric muon events. In order to compute the event spectra of the signal and the background contributions, we closely follow the approach of Ref.~\cite{Vincent:2016nut}, which in turn represents an update of the detailed calculations described in Ref.~\cite{Palomares-Ruiz:2015mka} (see also Refs.~\cite{Mena:2014sja, othersPalomares}), with some additional improvements. In this work, we use the angular and energy information of the spectra.

\begin{figure}[t]
	\begin{center}
		\includegraphics[width=0.76\textwidth]{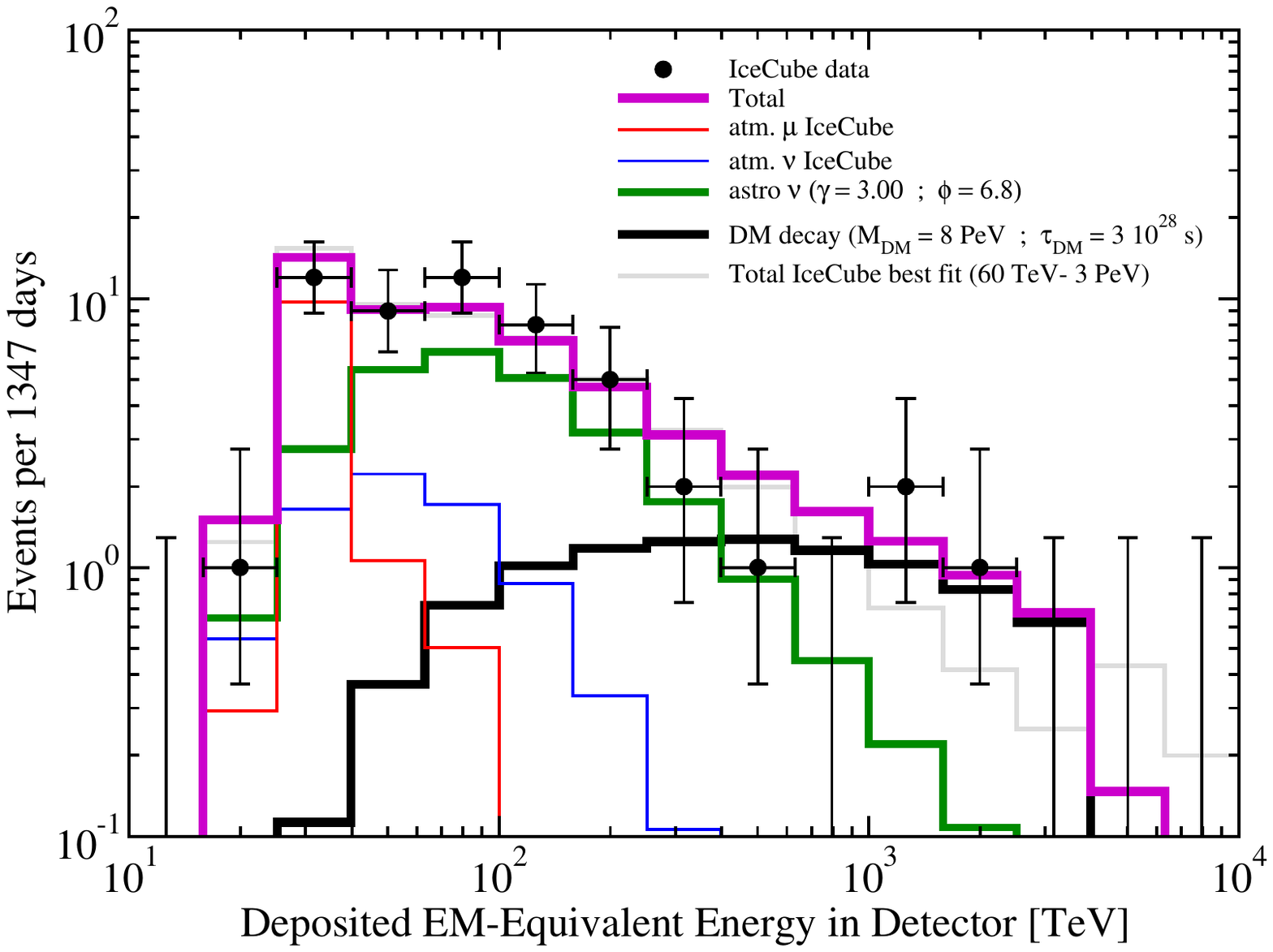} \vspace{-5mm} 
	\end{center}
	\caption{Same as Fig.~\ref{fig:events300}, but for an astrophysical spectrum $E_\nu^2 \, d\Phi_{\rm a}/dE_\nu = 6.8 \times 10^{-8} \, (E_\nu/100 \, {\rm TeV})^{-1} \, {\rm GeV} \, {\rm cm}^{-2} \, {\rm s}^{-1} \, {\rm sr}^{-1}$ and a DM signal for $M_{\rm DM} = 8$~PeV and $\tau_{\rm DM} = 3 \times 10^{28}$~s.}
	\label{fig:events8000}
\end{figure}

The event spectra for two cases: $(M_{\rm DM},  \tau_{\rm DM}) = (300~{\rm TeV}, 10^{28}~{\rm s})$; $(\gamma, \phi) = (2, 1.6)$ and $(M_{\rm DM},  \tau_{\rm DM}) = (8~{\rm PeV}, 3 \times 10^{28}~{\rm s})$; $(\gamma, \phi) = (3, 6.8)$, are shown (black histograms for the flux from DM decays and green histograms for the astrophysical flux), in Fig.~\ref{fig:events300} and Fig.~\ref{fig:events8000}, respectively. In both figures, we also show the background event spectra (red and blue histograms) and the total expected event spectra (purple histogram), along with the spectrum obtained using the preliminary IceCube best fit for $(1:1:1)_\oplus$ in the EM-equivalent deposited energy interval [60~TeV $-$ 3~PeV] (gray histogram) and the 4-year data points~\cite{Aartsen:2015zva}. Note that, for these values of the DM lifetime, the diffuse $\gamma-$ray background is well above the expected DM signal~\cite{Murase:2012xs, murase, Esmaili:2015xpa}.

In Fig.~\ref{fig:events300} we see that the DM signal (for $M_{\rm DM} = 300$~TeV and $\tau_{\rm DM} = 10^{28}~{\rm s}$) represents the dominant contribution between 40~TeV and 150~TeV, whereas the hard astrophysical power-law flux ($E_\nu^2 \, d\Phi_{\rm a}/dE_\nu = 1.6 \times 10^{-8} \, {\rm GeV} \, {\rm cm}^{-2} \, {\rm s}^{-1} \, {\rm sr}^{-1}$) would explain the high-energy part of the observed event spectrum. The small low-energy excess of events with respect to the 3-year results can be nicely explained by neutrinos from DM decays within the scenario described in this paper. This also implies that the astrophysical neutrino flux does not have to be as soft as the result of the fit with only such a flux~\cite{Aartsen:2015zva, Vincent:2016nut}. On the other hand, this hard spectrum is in agreement with the results obtained for the through-going muon sample~\cite{icecube2, Aartsen:2015knd}, with a per-flavour normalization which is slightly lower, yet compatible within errors. Moreover, let us also note that the through-going muon sample is sensitive mainly to energies from a few 100~TeV to a few PeV, which are precisely the energies in which the astrophysical flux in Fig.~\ref{fig:events300} is the dominant one. In addition, this hard astrophysical spectrum would not overshoot the $\gamma$-ray cascade limit~\cite{Murase:2013rfa, Chang:2014hua, Ando:2015bva}, or the data from air-showers arrays in galactic cases~\cite{Ahlers:2013xia, Kalashev:2014vra}, if $pp$ sources (where neutrinos are mainly produced from pion decays) are to explain this flux.

Finally, in Fig.~\ref{fig:events8000} we show the event spectra for a heavier DM candidate in combination with a softer power-law flux. In this case, the low-energy events can be explained by the astrophysical flux ($E_\nu^2 \, d\Phi_{\rm a}/dE_\nu = 6.8 \times 10^{-8} \, (E_\nu/100 \, {\rm TeV})^{-1} \, {\rm GeV} \, {\rm cm}^{-2} \, {\rm s}^{-1} \, {\rm sr}^{-1}$), whereas the prediction of the hard DM decay signal (for $M_{\rm DM} = 8$~PeV and $\tau_{\rm DM} = 3 \times 10^{28}~{\rm s}$) is in agreement with the highest-energy data. However, notice that the (almost) monochromatic flux of hard neutrinos does not translate into a bump in the total event energy spectrum. This is partly due to the particular flavor composition we chose, $f_{e, \oplus} \simeq 0.2$. On the other hand, the natural kinematical cutoff in the event spectrum from DM decays  (4~PeV in this case) could also explain the non-observation of events around the Glashow resonance energy ($E_\nu \sim 6.3$~PeV) and, in this case, the through-going muon data could be explained by the hard spectrum from DM decays\footnote{However, it is not possible to explain in this way the recently announced through-going muon event with deposited energy of 2.6~PeV~\cite{2.6PeV}, which is most likely produced by a $\sim$10~PeV muon neutrino (or a tau neutrino with higher energy)~\cite{Kistler:2016ask, Aartsen:2016xlq}, unless one considers a much heavier DM candidate. Note, as well, that the kinematical cutoff would not solve the current tension between the lack of Glashow events and the observation of this very energetic through-going muon track.}, instead. Let us finally note that we have not shown the results for lighter $M_{\rm DM} \simeq \cal{O}(\rm PeV)$, usually quoted in the literature~\cite{serpico, Esmaili:2014rma, Fong:2014bsa, Daikoku:2015vsa} when considering earlier IceCube data. However, it does not represent a better agreement with data than the case shown in Fig.~\ref{fig:events8000}. Given the soft event spectrum resulting from the 4-year IceCube HESE data, decays from such a DM candidate, cannot explain the entire observed event spectra anymore.

\section{Conclusions and final remarks}
\label{sec:conclusions}

As we discussed in detail, in the scenario of cold DM from RH neutrino mixing, the same new interactions are responsible both for $N_{\rm DM}$ production and DM decays, with much stronger predictive power compared to models one can imagine where there is  one kind of interaction responsible for production and another responsible for decays (e.g., some tiny small Yukawa coupling) where one has in any case freedom to reproduce both DM abundance and a contribution to IceCube neutrinos. 

Therefore, finding viable solutions which are able to accommodate leptogenesis, a good DM candidate and are testable signal with neutrino telescopes is highly non-trivial. This is thanks to the possibility to generate the $N_{\rm DM}$ abundance when the $N_{\rm S}$ is still not fully thermalised, an observation that makes viable the hierarchical case with $M_{\rm DM}\gtrsim M_{\rm S}$. Physically, this relaxes the bounds since $N_{\rm S}$ can be light with a small Yukawa coupling for higher $T_{\rm res}$, both things helping DM stability (see Eqs.~(\ref{tau2bodybis}) and~(\ref{rate4body})) and efficiency of production (see Eq.~(\ref{gres})). 

In this way, even starting from initial vanishing $N_{\rm S}$ abundance, there is an allowed range for the DM mass that, depending on the ratio $M_{\rm DM}/M_{\rm S}$, extends from $\sim 100$~GeV to about $\sim 10$~PeV (for $\xi \lesssim 10$). On the other hand, the higher the value of $\t_{\rm DM}^{\rm min}$, the narrower the allowed range of masses. For instance from the Fig.~\ref{fig:bounds} one can see that for $\tau_{\rm DM}^{\rm min} \gg 10^{29}$~s, the case with $M_{\rm DM}/M_{\rm S}=10$ and $\xi=1$ would be ruled out. More generally the upper bound for $M_{\rm DM}$ would become more and more stringent. In addition, the existence of a DM candidate nicely combines with a two-RH neutrino scenario of (resonant) leptogenesis to successfully reproduce the correct baryon asymmetry at a scale below $\sim 10$~TeV (1~PeV) for initial vanishing (thermal) $N_{\rm S}$ abundance. In this case, the allowed range of values for  $M_{\rm DM}$ narrows to $\sim 1$~TeV --- 300~TeV, for $\xi \lesssim 10$, 
in order to have $T_{\rm lep}>T_{\rm sph}^{\rm out}$ so that a lepton asymmetry can be reprocessed into a baryon asymmetry.  

A contribution from $N_{\rm DM}$ decays to the high-energy neutrino flux can help reproducing the IceCube data in addition to an astrophysical component that is, in any case, necessary. Without performing a dedicated fit to the data, we have shown the contribution from DM decays to the energy spectrum for two exemplary masses, $M_{\rm DM}=300$~TeV and $M_{\rm DM}=8$~PeV, that could help explaining some of the features in the current HESE data. However, we have not investigated which case is statistically preferred (see Ref.~\cite{BEPS}). Nevertheless, we do not find that a mass of $M_{\rm DM}\simeq 4$~PeV, discussed in the literature, is particularly favoured with the current data and definitely it cannot explain the entire event spectrum.

 In principle, neutrinos produced directly from the decays of $N_{\rm DM}$ via $N_{\rm S}$ retain information on the Yukawa couplings and might be singled out from the rest at energies close to $M_{\rm DM}/2$. Albeit challenging, this is an interesting possibility, as the flavour composition typically differs from conventional astrophysical flavour ratios. During the next years it will be interesting to see whether more high-energy neutrino events are detected, which could support (depending on the flavor composition) the presence of a component originating from the decays of DM RH neutrinos produced by mixing with source RH neutrinos.

\section*{Acknowledgments}

This paper is dedicated to the memory of Alexey Anisimov. Our work relies on many of his ideas and results. We also wish to thank Bhupal Dev, Arman Esmaili, Steve King, Stefano Morisi, Nobuchilka Okada and Apostolos Pilaftis for useful comments. PDB acknowledges financial support from the NExT/SEPnet Institute.  PDB is also grateful to the Mainz Institute for Theoretical Physics (MITP) for its hospitality and its partial support during the completion of this work.  PDB and PL acknowledge financial support from the STFC Consolidated Grant ST/L000296/1. SPR is supported by a Ram\'on y Cajal contract, by the Spanish MINECO under grants FPA2014-54459-P and SEV-2014-0398 and by the Generalitat Valenciana under grant PROMETEOII/2014/049. This project has also received funding from the European Union's Horizon 2020 research and innovation programme under the Marie Sklodowska-Curie grant agreement No 690575 and No 674896. SPR is also partially supported by the Portuguese FCT through the CFTP-FCT Unit 777 (PEst-OE/FIS/UI0777/2013).

\end{document}